\documentclass[journal=ancac3,manuscript=article]{achemso}

\usepackage[version=3]{mhchem} 
\usepackage[T1]{fontenc}       
\usepackage{graphicx}
\usepackage{dcolumn}
\usepackage{float}
\usepackage{amsmath}
\usepackage{amssymb}
\usepackage{verbatim}

\author{Ignacio Gimeno}
\affiliation[ICMA]{Instituto de Ciencia de Materiales de Arag\'on, CSIC-Universidad de Zaragoza, Pedro Cerbuna 12, 50009 Zaragoza, Spain}
\author{Wenzel Kersten}
\affiliation[TUW]{Vienna Center for Quantum Science and Technology, Atominstitut, TU Wien, 1020 Vienna, Austria}
\author{Mar\'{\i}a C. Pallar\'es}
\author{Pablo Hermosilla}
\affiliation[INA]{Laboratorio de Microscop\'{\i}as Avanzadas, Instituto de Nanociencia de Arag\'on, Universidad de Zaragoza, 50018 Zaragoza, Spain}
\author{Mar\'{\i}a Jos\'e Mart\'{\i}nez-P\'erez}
\affiliation[ICMA]{Instituto de Ciencia de Materiales de Arag\'on, CSIC-Universidad de Zaragoza, Pedro Cerbuna 12, 50009 Zaragoza, Spain}
\alsoaffiliation{Fundaci\'on ARAID, Campus Río Ebro, 50018 Zaragoza, Spain}
\author{Mark D. Jenkins}
\affiliation[ICMA]{Instituto de Ciencia de Materiales de Arag\'on, CSIC-Universidad de Zaragoza, Pedro Cerbuna 12, 50009 Zaragoza, Spain}
\author{Andreas Angerer}
\affiliation[TUW]{Vienna Center for Quantum Science and Technology, Atominstitut, TU Wien, 1020 Vienna, Austria}
\author{Carlos S\'anchez-Azqueta}
\affiliation[UZaragoza]{Departamento de F\'{\i}sica Aplicada, Universidad de Zaragoza, 50009 Zaragoza, Spain}
\author{David Zueco}
\affiliation[ICMA]{Instituto de Ciencia de Materiales de Arag\'on, CSIC-Universidad de Zaragoza, Pedro Cerbuna 12, 50009 Zaragoza, Spain}
\alsoaffiliation{Fundaci\'on ARAID, Campus Río Ebro, 50018 Zaragoza, Spain}
\author{Johannes Majer}
\affiliation[USTC]{Shanghai Branch, CAS Center for Excellence and Synergetic Innovation Center in Quantum Information and Quantum Physics, University of Science and Technology of China, Shanghai 201315, China}
\alsoaffiliation{National Laboratory for Physical Sciences at Microscale and Department of Modern Physics, University of Science and Technology of China, Hefei 230026, China}
\altaffiliation{Vienna Center for Quantum Science and Technology, Atominstitut, TU Wien, 1020 Vienna, Austria}
\author{Anabel Lostao}
\affiliation[ICMA]{Instituto de Ciencia de Materiales de Arag\'on, CSIC-Universidad de Zaragoza, Pedro Cerbuna 12, 50009 Zaragoza, Spain}
\alsoaffiliation[INA]{Laboratorio de Microscop\'{\i}as Avanzadas, Instituto de Nanociencia de Arag\'on, Universidad de Zaragoza, 50018 Zaragoza, Spain}
\alsoaffiliation{Fundaci\'on ARAID, Campus Río Ebro, 50018 Zaragoza, Spain}
\author{Fernando Luis}
\email{fluis@unizar.es}
\affiliation[ICMA]{Instituto de Ciencia de Materiales de Arag\'on, CSIC-Universidad de Zaragoza, Pedro Cerbuna 12, 50009 Zaragoza, Spain}
\title {Enhanced molecular spin-photon coupling at
superconducting nanoconstrictions}


\begin{document}

\begin{tocentry}

\centering
\includegraphics[width=9 cm]{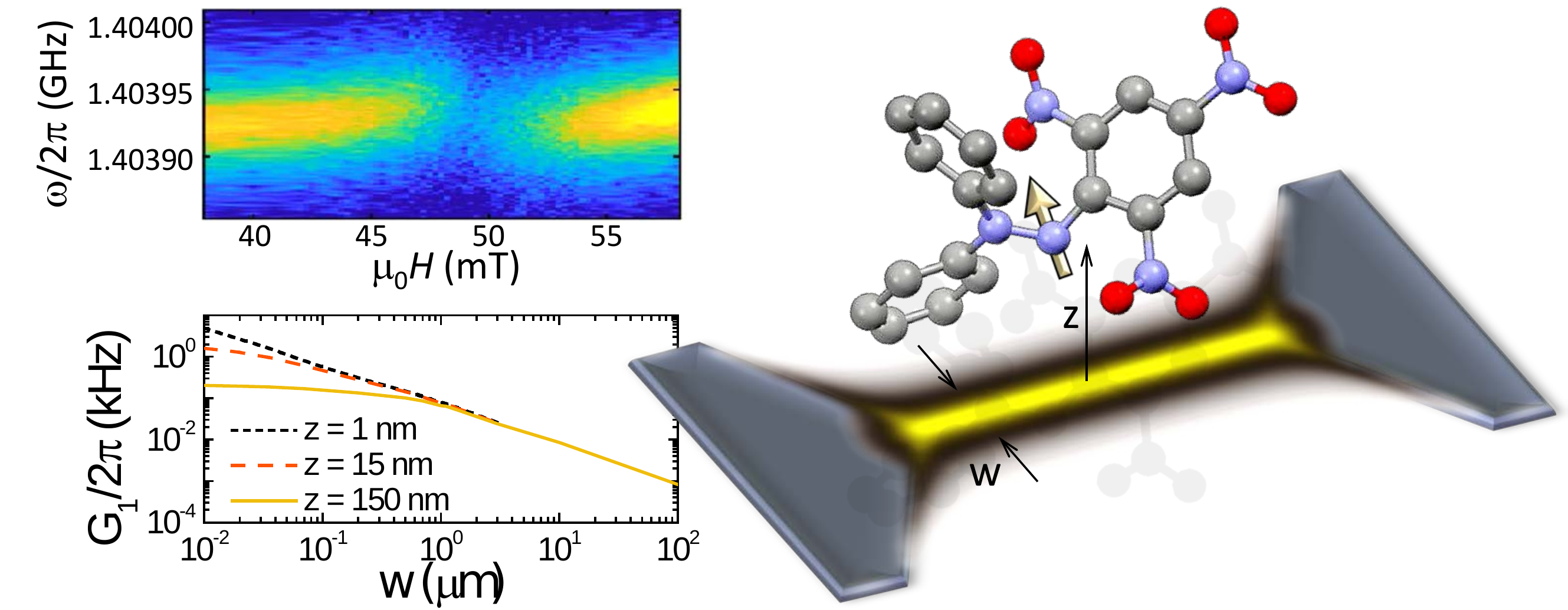}

Top-down and bottom-up nanolithographies are combined to show that the coupling of molecular spins to microwave circuits is strongly enhanced when the former are in close proximity of a superconducting nanoconstriction





\end{tocentry}

\begin{abstract}
We combine top-down and bottom-up nanolithography to optimize the coupling of small molecular spin ensembles to $1.4$ GHz on-chip superconducting resonators. Nanoscopic constrictions, fabricated with a focused ion beam at the central transmission line, locally concentrate the microwave magnetic field. Drops of free-radical molecules have been deposited from solution onto the circuits. For the smallest ones, the molecules were delivered at the relevant circuit areas by means of an atomic force microscope. The number of spins $N_{\rm eff}$ effectively coupled to each device was accurately determined combining Scanning Electron and Atomic Force Microscopies. The collective spin-photon coupling constant has been determined for samples with $N_{\rm eff}$ ranging between $2 \times 10^{6}$ and $10^{12}$ spins, and for temperatures down to $44$ mK. The results show the well-known collective enhancement of the coupling proportional to the square root of $N_{\rm eff}$. The average coupling of individual spins is enhanced by more than four orders of magnitude (from $4$ mHz up to above $180$ Hz) when the transmission line width is reduced from $400$ microns down to $42$ nm, and reaches maximum values near $1$ kHz for molecules located on the smallest nanoconstrictions. This result opens promising avenues for the realization of magnetic spectroscopy experiments at the nanoscale and for the development of hybrid quantum computation architectures based on molecular spin qubits.
\end{abstract}


The coupling of spins to superconducting circuits lies at the basis of diverse technologies. Superconducting on-chip resonators, which concentrate the microwave magnetic field in much smaller regions than conventional three-dimensional cavities,\cite{Wallraff2004,Goppl2008} promise to take electron spin resonance (ESR) to its utmost sensitivity level, eventually allowing the detection of single spins.\cite{Bienfait2016,Eichler2017,Probst2017,Sarabi2019} Besides, single microwave photons "trapped" in these devices provide a way to wire-up qubits,\cite{Blais2004,Majer2007,Schoelkopf2008} and therefore form a basis for hybrid quantum computation and simulation schemes based on spins.\cite{Imamoglu2009,Wesenberg2009,Schuster2010,Kubo2010,Wu2010,Chiorescu2010,Jenkins2016}

Different approaches have been designed and, in some cases, put into practice, to enhance the spin sensitivity and the spin-photon coupling. They often involve the use of nonlinear superconducting circuits, parametric amplifiers, to amplify the output signal,\cite{Bienfait2016,Eichler2017} and of low impedance resonator designs, which increase the photon magnetic field.\cite{Eichler2017,Probst2017} Experiments performed on highly coherent magnetic impurities in semiconducting hosts show spin resonance at the level of a few tens of spins and maximum spin-photon coupling strengths of order $400$ Hz at frequencies of about $7$ GHz.\cite{Probst2017}

In this work, we explore experimentally a third alternative, inspired by earlier developments of micro-ESR devices.\cite{Narkowicz2005,Narkowicz2008,Banholzer2011} The underlying idea is that reducing the cavity effective volume enhances the microwave magnetic field. In a superconducting resonator, this goal can be achieved by decreasing locally the width of the resonator's transmission line \cite{Jenkins2013,Jenkins2014,Haikka2017} in order to bridge the very different scale lengths of superconducting circuits, with typical line widths of a few microns, and of impurity or molecular spins, which range in the range of nanometers. This approach requires moving beyond the limits of conventional optical lithography to fabricate or modify certain regions of the circuits and is fully complementary to those methods mentioned above. 

The main challenge resides in the fact that the microwave field enhancement is localized in a nanoscopic region near the superconducting constriction, as can be seen in Figs. \ref{fgr:circuit}A and \ref{fgr:circuit}B. Optimally profiting from such enhancement thus calls for a method able to deliver the magnetic sample at the right location and with sufficient spatial accuracy. Spins in molecules provide a good test case for addressing this challenge. Many molecules are stable in solution or can be sublimated from the crystal, and can therefore be transferred to a solid substrate or a device.\cite{Mannini2009,Mannini2010,Domingo2012,Thiele2014,Malavolti2018,Urtizberea2020} In addition, they are one of the most promising candidates to encode spin qubits, due to the vast possibilities for design offered by chemical nanoscience.\cite{Leuenberger2001,Troiani2011,Aromi2012,Moreno-Pineda2018,Gaita2019,Atzori2019} Here, we combine top-down and bottom-up nanolithography methods to study the coupling of microwave photons to nanoensembles of the simplest molecular spins, organic free radicals with $S=1/2$, deposited near constrictions of varying size.         

\section{Results and discussion}
This section describes the main results of this work: how devices are modified to locally enhance the spin-photon coupling, how the molecules are deposited with sufficient spatial accuracy, and how this coupling is experimentally determined as a function of the number of molecules and of temperature. The main result refers to the coupling of individual spins and its dependence on the size of the resonator dimensions and on the molecule to chip interface.

\subsection{Circuit fabrication and integration of molecular spin micro- and nano-deposits}

\begin{figure}
\centering
\includegraphics[width=0.99\columnwidth]{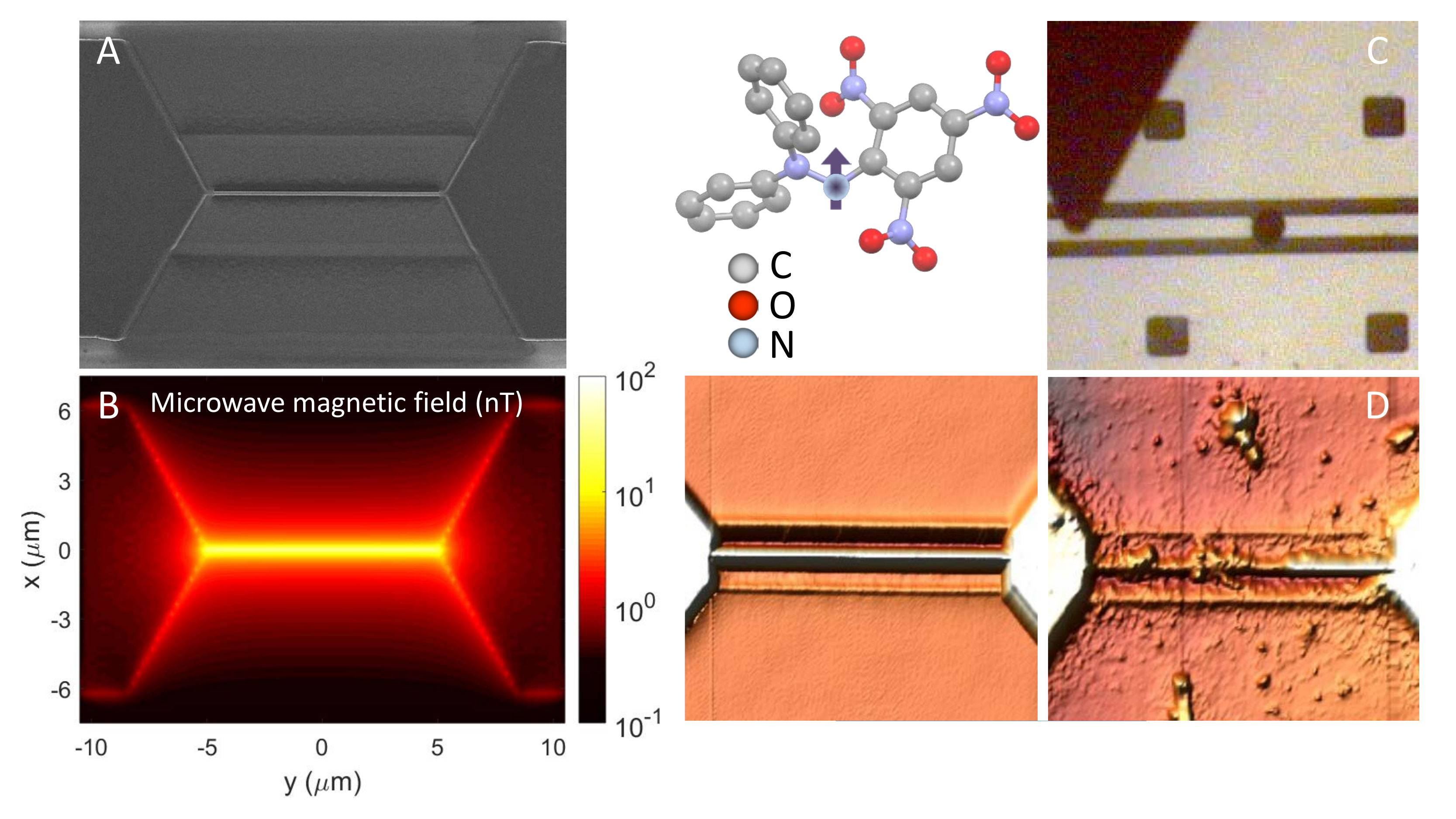}
  \caption{ A: Scanning electron microscopy image of the central line of a superconducting coplanar resonator. The line was thinned down to a width of about $158$ nm by focused ion beam nano-lithography. B: Color plot of the photon magnetic field in the neighborhood of this constriction, calculated with a finite-element simulation software.\cite{Khapaev2002} C: Structure of a DPPH free-radical molecule, with spin $S=1/2$ and $g=2$ (left) and optical microscopy image of the constriction after the deposition of DPPH by means of the tip of an Atomic Force Microscope(AFM, right). D: AFM image of the constriction taken before and after the molecules were deposited and the solvent had evaporated.}
  \label{fgr:circuit}
\end{figure}

The circuits used in this work are coplanar superconducting resonators fabricated by optical lithography on $150$ nm thick Nb films, which become superconducting below $T_{\rm c} = 8.2$ K, deposited onto single crystalline sapphire wafers. Previous studies \cite{Jenkins2014} show that nanoscopic constrictions, such as the one shown in Fig. \ref{fgr:circuit}A, can be fabricated in the central transmission line by Focused Ion Beam (FIB) nanolithography, that they do not significantly alter the operation of the devices and that they locally enhance the photon magnetic field. 

As the simulation in Fig. \ref{fgr:circuit}B shows, the enhancement is strongly localized near the superconducting nanobridge, in a region with typical dimensions $w \times w \times L$, where $w$ is the constriction width and $L$ its length. The proper integration of spins plays, then, a crucial role to achieve the maximum spin sensitivity allowed by this approach. Our samples contain organic free-radical molecules of $2$,$2$-diphenyl-$1$-picrylhydrazyl,\cite{Weil1965,Yordanov1996,Zilic2010} hereafter referred to as DPPH, whose molecular structure is shown in Fig. \ref{fgr:circuit}. Each molecule hosts an unpaired electron with a spin $S=1/2$ and a close to isotropic gyromagnetic factor $g \simeq 2$. Under a magnetic field $H$, it shows a well-defined resonance line at a frequency $\Omega_{S} = g \mu_{\rm B} H/\hbar$. In a crystalline environment, the inhomogeneous broadening arising from dipolar interactions is reduced by direct exchange interactions between nearest radicals.\cite{Anderson1953,Hocherl1965} The resonance linewidth becomes then dominated by the homogeneous broadening $1/{\rm T}_{2}$, where $T_{2} \simeq 80-120$ ns is the spin coherence time. Besides, these molecules can be dissolved and remain stable in diverse organic solvents.\cite{Yordanov1996}

The latter property allows delivering precise amounts of DPPH molecules onto the central transmission lines of the resonators. Large (a few mm wide, a few microns thick) molecular ensembles have been deposited either in powder form or from solution, using a micropipette, onto conventional resonators with $w = 400\mu$m and $14 \mu$m. The number of molecules was varied by controlling the concentration of the original solution, keeping the volume constant. Illustrative Scanning Electron Microscopy (SEM) images of such deposits are shown as part of the supplementary information. They show that DPPH has a tendency to form quasi-spherical nano-aggregates. 

For smaller deposits on narrower lines, we employed Dip Pen Nanolithography (DPN),\cite{Piner1999,Bellido2010} a soft lithographic technique that uses the tip of an Atomic Force Microscope (AFM) to deposit nanoscopic volumes from a solution containing the molecules of interest onto a very small area (cf Fig. \ref{fgr:circuit}C). This technique has the advantage of combining high spatial resolution with a good control over the molecular dose transferred to the substrate without the need of chemical functionalization, thus making it well suited for placing diverse nano-samples onto solid state sensors.\cite{Martinez2011,Domingo2012,Bellido2013} The size of the deposits was controlled in this case by the diameter of the drops transferred by the AFM tip to the substrate, which depend on the contact time between both, and by their concentration. The deposits are then characterized by SEM and by AFM, as shown in Figs. \ref{fgr:circuit}D and in the supplementary information. These deposits also form nano-aggregates after the solvent has evaporated, with sizes ranging from $50$ nm to above $200$ nm depending on the initial concentration of the radical at the ink solution and the size of the deposited drop. This also means that the number of molecules actually transferred into the "active" region of the constriction might vary even between depositions performed under nominally identical conditions and, therefore, needs to be determined for each case. 

\subsection{Spin-photon coupling versus spin number}

The coupling of the spins to the cavity photons, with frequency $\omega_{\rm r}/2 \pi \simeq 1.4$ GHz, reduces the microwave transmission, shifts $\omega_{\rm r}$ and broadens the resonance. All these effects become maximum when $\Omega_{S} \simeq \omega_{\rm r}$, i.e. when spins and photons are brought into resonance with each other by an external magnetic field $H$. The transmission provides information on the collective coupling $G_{N}$ of the spin ensemble to the resonator, where $N$ denotes here the number of free-radical spins, as well as on the spin $\gamma$ and photons $\kappa$ characteristic line widths.   

\begin{figure}
\centering
\includegraphics[width=0.99\columnwidth]{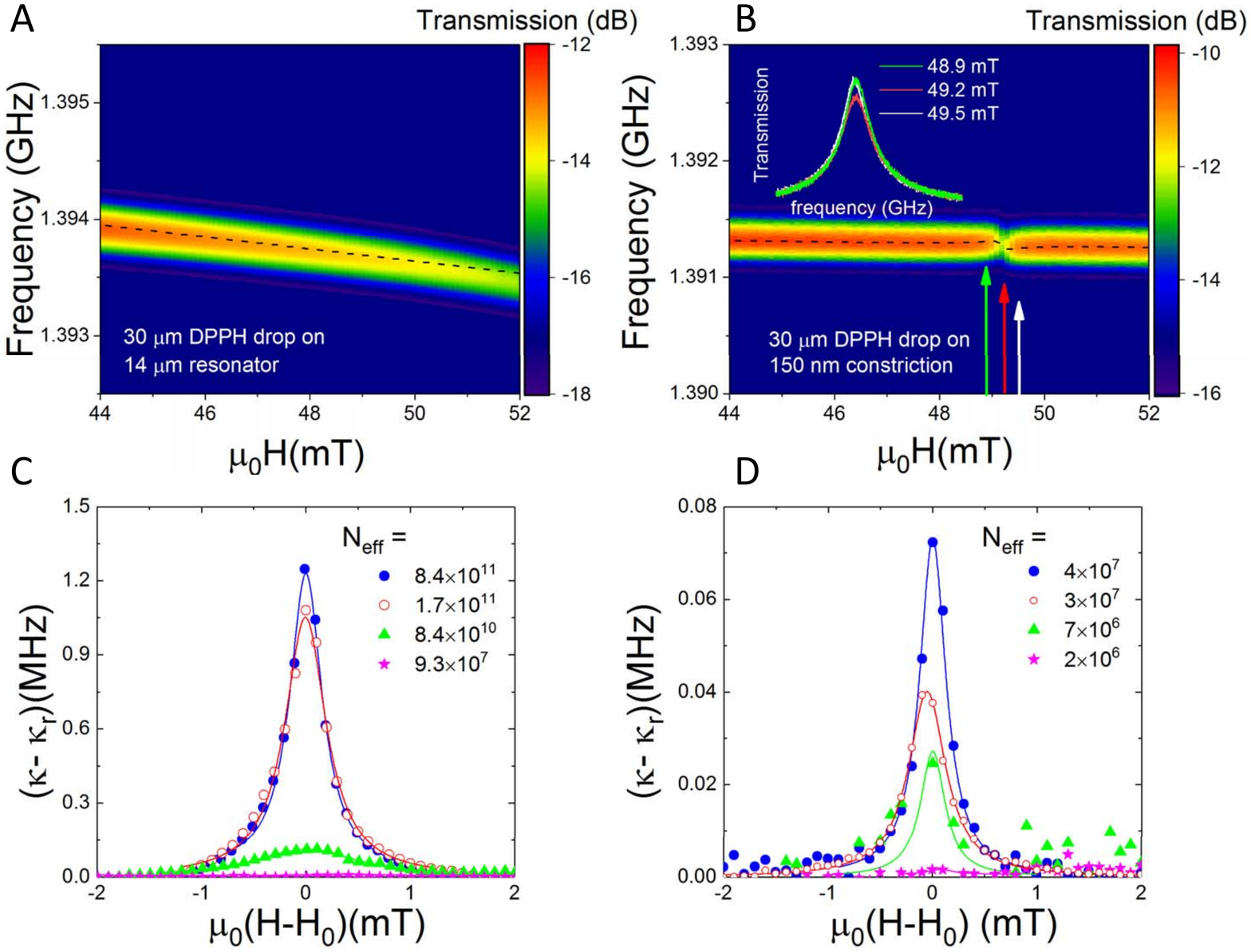}
  \caption{ Top: Color scale plots of the microwave transmission through $1.4$ GHz on-chip superconducting resonators with a $14 \mu$m wide central line (A) and with a $158$ nm wide constriction (B) coupled to a DPPH deposit with $N \sim 5 \times 10^{9}$ molecules, corresponding to $N_{\rm eff} \sim 4 \times 10^{7}$ spins effectively coupled to the resonator at $T=4.2$ K. The red dashed lines mark the position of the resonance frequency at each magnetic field. The inset in B shows how the resonance in the latter device evolves with magnetic field (at the field values indicated by arrows), evidencing the detection of a net absorption (lower transmission and broader resonance) when the spins get into mutual resonance with the circuit. Bottom: Magnetic field dependence of the resonance width $\kappa$ for the same resonators (C without and D with constriction) coupled to ensembles of DPPH molecules of varying size. Solid lines are least square fits based on Eq. (\ref{eqn:kappa}).}
  \label{fgr:2Dplots}
\end{figure}

Illustrative results of transmission measurements performed at $T=4.2$ K on resonators without and with a $158$ nm wide constriction are shown in Fig. \ref{fgr:2Dplots}A and \ref{fgr:2Dplots}B, respectively. In both cases, the sample was a DPPH drop with an approximate diameter of about $30 \mu$m deposited onto the central line by DPN. This amount corresponds approximately to the sensitivity limit for the former device. By contrast, under the same conditions, the resonator with the constriction gives a clearly visible absorption signal, thus providing direct evidence for the enhancement of $G_{N}$. 

In order to obtain quantitative estimates of this enhancement, and estimate the average coupling to individual spins, experiments on samples with decreasing $N$ were performed. The number of molecules effectively coupling to the resonator magnetic field was estimated from the geometry and topography of the deposits, taking into account the width of the transmission line near the deposit. For conventional resonators with a $14 \mu$m wide central line, all spins located in a $30 \mu$m wide region around it are counted. For a nanoconstriction, the "active" area is taken as a $2 \mu$m wide rectangular area around it. Experiments and simulations, described below and in the supplementary information, confirm that the coupling of spins located outside this region lies below the sensitivity limits and can, therefore, be safely neglected. Another aspect that needs to be taken into account is that these measurements were performed at a finite temperature $T=4.2$ K. This reduces the population difference, or equivalently the spin polarization $\langle S_{z} \rangle_{T}$, between the ground and excited levels of the DPPH spins by a factor $6.7 \times 10^{-3}$ with respect to zero temperature. This also reduces the number of spins that contribute to the net absorption, from $N$ to $N_{\rm eff} = N \langle S_{z} \rangle_{T}/S$, where $\langle S_{z} \rangle_{T}/S = \tanh{g \mu_{\rm B} H S/k_{\rm B} T}$ is the spin polarization achieved by the magnetic field at the given temperature.   

Results from these experiments are shown in Figs. \ref{fgr:2Dplots}C and \ref{fgr:2Dplots}D. They confirm that introducing a nanoconstriction enables detecting much smaller deposits: the sensitivity limit is reduced from about $10^{8}$ spins to $2 \times 10^{6}$ spins. This enhancement in sensitivity arises here from a larger spin-photon coupling, which can be determined as follows. The broadening of the cavity resonance $\kappa$ can be fitted using the following expression,\cite{Bushev2011}

\begin{equation}
\kappa = \kappa_{\rm r} + \frac{G_{N}^{2}\gamma}{\left(\omega_{\rm r} - \Omega_{S} \right)^{2} + \gamma^{2}} 
   \label{eqn:kappa}
\end{equation}

\noindent where $\kappa_{\rm r}$ is the broadening of the 'empty' cavity, as measured when it is detuned from the spins. This fit allows extracting $G_{N}$ and $\gamma$. The latter is found to be close to $12$ MHz for all but the smallest DPN deposits. The former value is compatible with a pure homogeneous broadening and a ${\rm T}_{2} \simeq 80$ ns, which lies within the typical values observed for DPPH.\cite{Yordanov1996} A larger line width has been observed in DPPH molecules dispersed in polymeric matrices.\cite{Hocherl1965} That the same phenomenon is observed in DPPH nano-aggregates deposited from solution suggests that they tend to lose crystalline order. 

\begin{figure}
\centering
\includegraphics[width=0.5\columnwidth]{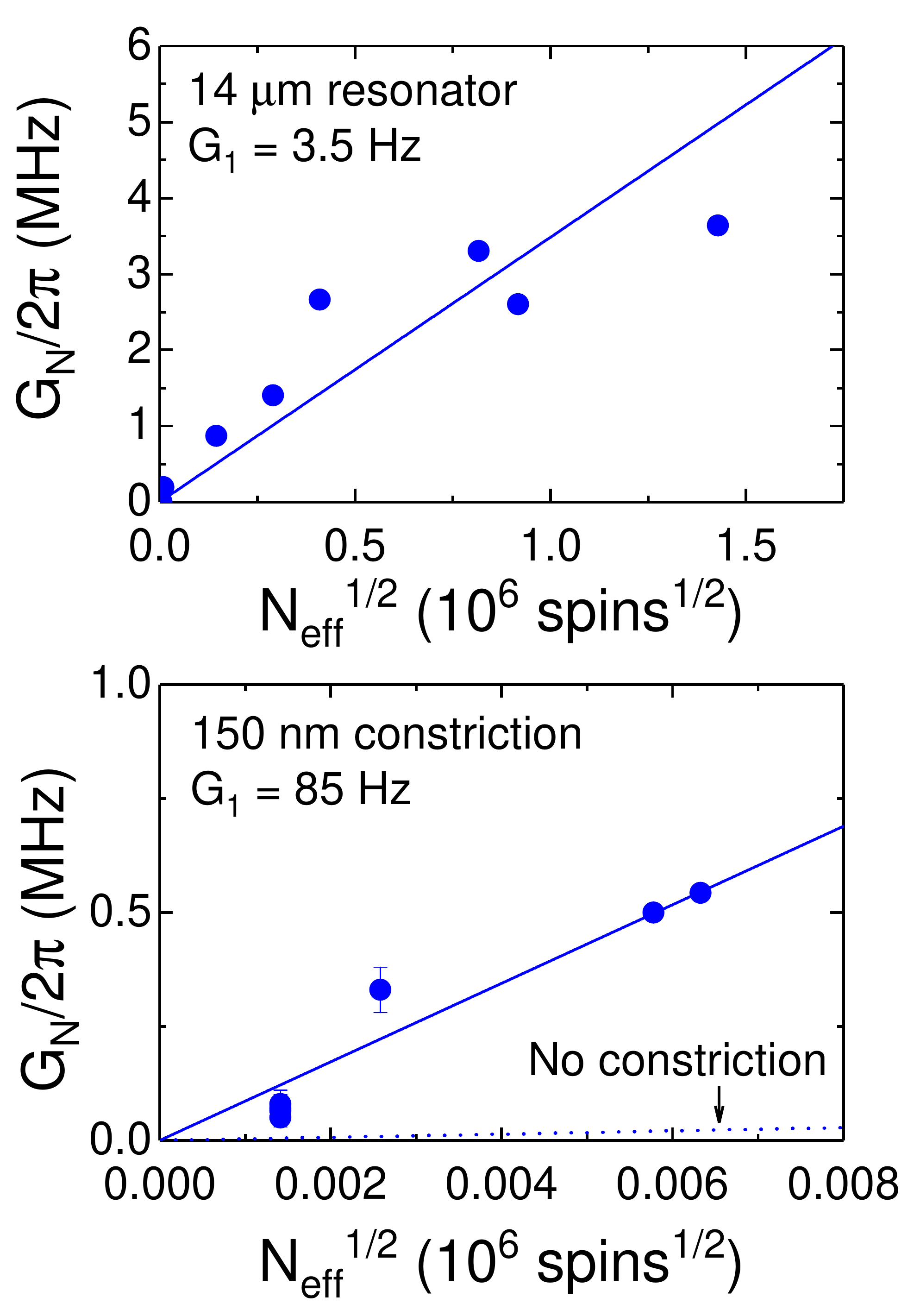}
  \caption{Collective spin-photon coupling of ensembles of free-radical molecules to coplanar resonators with a $14 \mu$m wide transmission line (top) and a $158$nm constriction (bottom). The solid lines are least-square fits to a linear dependence on the square root of the effective number of spins that are coupled to the devices at $T=4.2$ K, as predicted by Eq. (\ref{eqn:GN}). The bottom panel compares both fits to highlight the coupling enhancement generated by the constriction.}
  \label{fgr:gvsNeff}
\end{figure}

Figure \ref{fgr:gvsNeff} shows the dependence of $G_{N}$  on $N_{\rm eff}$ measured in resonators with $14 \mu$m and $158$ nm transmission line widths. In both, the coupling is approximately proportional to $N_{\rm eff}^{1/2}$. This result agrees with the collective enhancement of the radiation emission and absorption by spin ensembles,\cite{Tavis1968} which predicts that

\begin{equation}
G_{N} = G_{1} N_{\rm eff}^{1/2} 
   \label{eqn:GN}
\end{equation}

\noindent where $G_{1}$ is the coupling of a single spin. The  ability to modify the size of the molecular ensembles over a large range allows monitoring this well-known dependence directly. The slope then directly gives the average, or typical, value of $G_{1}$. These data confirm also that a strong enhancement of $G_{1}$, by a factor of order $24$ (from  $G_{1} / 2 \pi \simeq 3.5$ Hz to $85$ Hz), occurs when $w$ is reduced by a factor $100$ (from $14 \mu$m to $158$ nm). 

\subsection{Spin-photon coupling versus temperature: coupling of very small spin ensembles in the quantum regime}

In this section, we describe experiments that explore the optimal conditions in the search of a maximum spin-photon coupling: temperatures close to absolute zero, which take $N_{\rm eff}$ closer to $N$, an average number $n_{\rm photons} \simeq 5 \times 10^{5}$ of photons in the cavity well below $N$, in order to avoid any saturation effects, and a central line width $w \simeq 42$ nm near the minimum achievable by FIB nanolithography. Representative images of the constriction and of the DPPH nano-aggregates deposited on it by DPN are shown in Fig. \ref{fgr:AFM} and in the supplementary material. The number $N$ of DPPH molecules in the active area was, in this case, estimated to be approximately $1.6 \times 10^{8}$ spins, which lies below the minimum dose that was detectable in experiments performed at $T=4.2$ K and for $w=158$ nm (cf Fig. \ref{fgr:gvsNeff}D). 

\begin{figure}
\centering
\includegraphics[width=0.99\columnwidth]{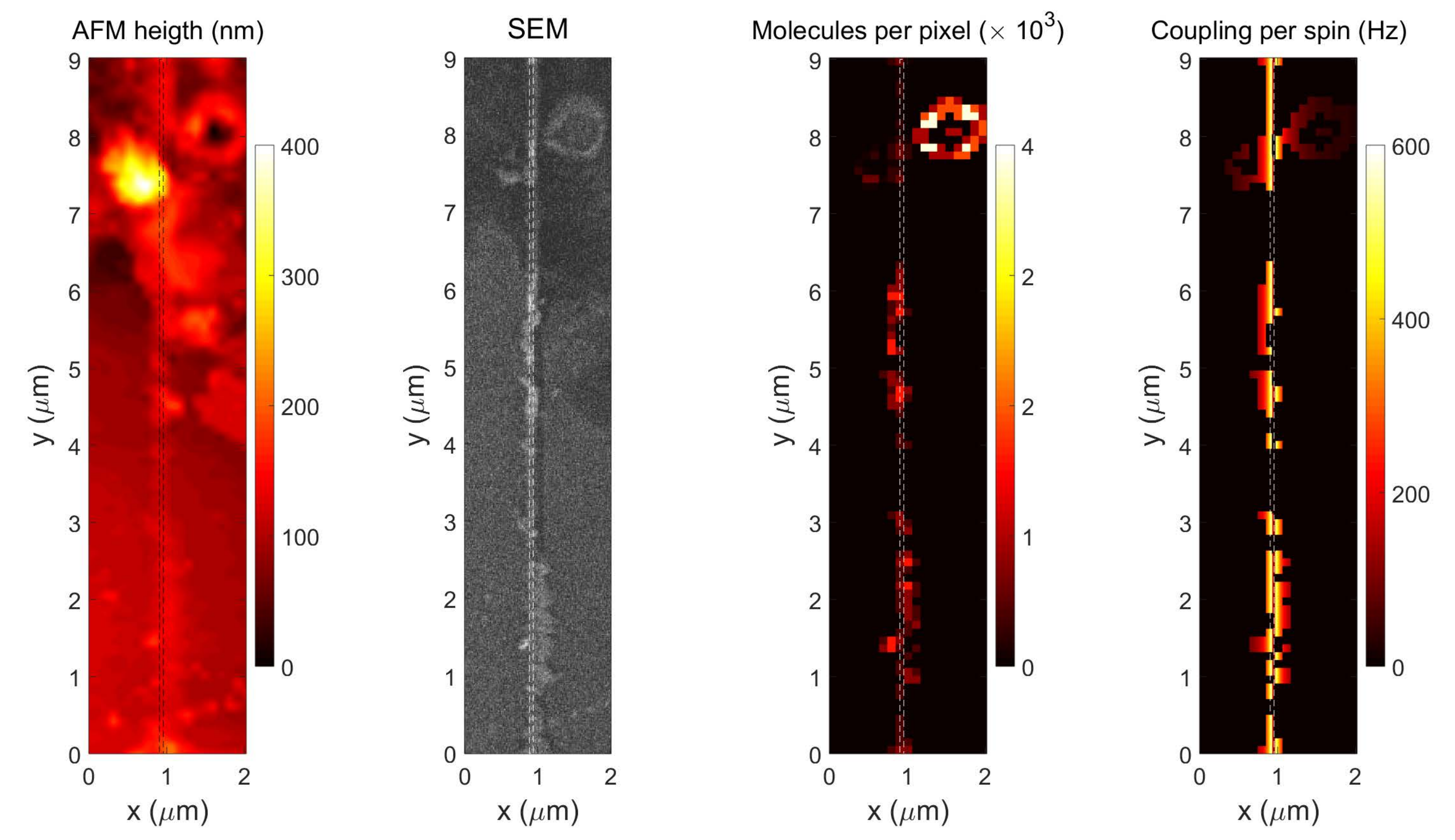}
  \caption{From left to right: AFM and SEM images of the region near a $42$ nm wide nano-constriction, color map of the number of DPPH molecules deposited on each location, and color map of the estimated single spin to photon couplings. The latter maps have been calculated with a discretization of space into $3 \times 3 \times 3$ nm$^{3}$ cubic cells. The number of free-radical molecules deposited in this area amounts to approximately $N = 1.6 \times 10^{8}$ and the collective coupling estimated from the simulations is $G_{N} / 2 \pi \simeq 2.0$ MHz at $T=44$ mK and $2.52$ MHz at $T=0$.}
  \label{fgr:AFM}
\end{figure}

\begin{figure}
\centering
\includegraphics[width=0.99\columnwidth]{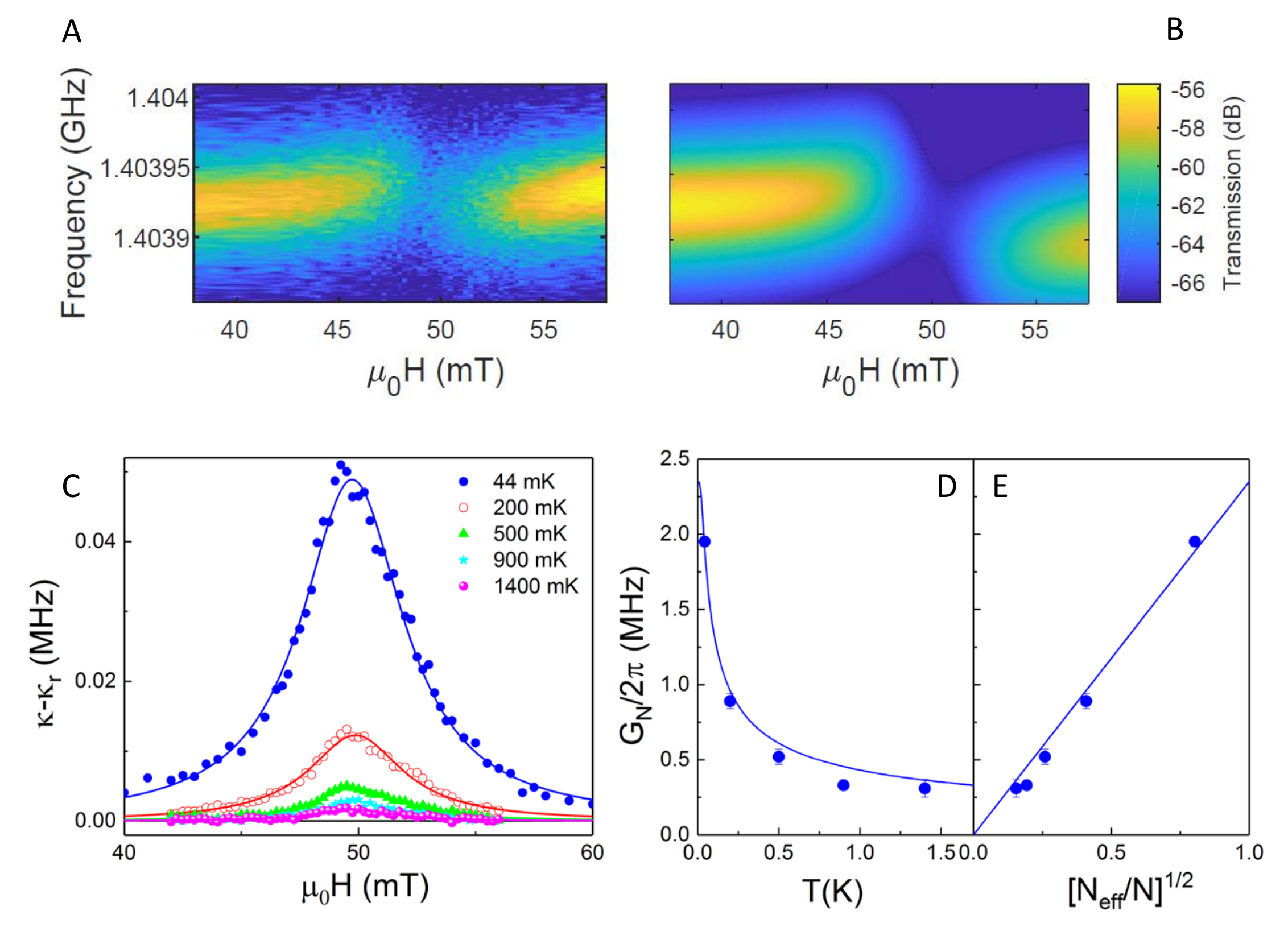}
  \caption{A: Color plot of the microwave transmission, measured at $T=44$mK, through a superconducting resonator with a $42$ nm wide constriction in its central transmission line and coupled to an ensemble of $N \simeq 1.6 \times 10^{8}$ free-radical molecules (corresponding to $N_{\rm eff} \simeq 10^{8}$ spins effectively coupled to the resonator). B: Color plot of the microwave transmission calculated for a collective coupling  $G_{N} / 2 \pi = 2.0$ MHz, as follows from the simulations described in Fig. \ref{fgr:AFM}, and a spin line width $\gamma = 65$ MHz. C: Magnetic field dependence of the resonance width $\kappa$ for the same device measured at different temperatures. Solid lines are least-square fits using Eq. (\ref{eqn:GN}). D: Temperature dependence of $G_{N}$ extracted from these experiments. E: Same data plotted as a function of the (temperature dependent) effective number of spins coupled to the resonator, which again show the linear dependence of $G_{N}$ on $\sqrt{N_{\rm eff}}$ and which extrapolate to $G_{N} / 2 \pi \simeq 2.32$ MHz for  $T \rightarrow 0$ (when $N_{\rm eff} \rightarrow N$).}
  \label{fgr:gvsT}
\end{figure}

Results of transmission measurements performed on this device are shown in Fig. \ref{fgr:gvsT}. Thanks to the much larger spin polarization, an easily discernible absorption signal is observed at the minimum temperature $T \simeq 44$ mK, which corresponds to a collective coupling $G_{N} / 2 \pi \simeq 1.9$ MHz. The linewidth $\gamma \simeq 65$ MHz is $5$ times larger than what would be expected from the spin coherence times. The additional broadening probably arises from the smaller size of the DPPH nano-aggregates transferred to this device and from the fact that DPPH molecules are here dispersed in a matrix of glycerol used in the DPN deposition process (see the section on methods below and the supplementary information for more details). As mentioned above, these effects tend to suppress direct exchange interactions between free-radical spins and, then, enhance the broadening associated with hyperfine and dipole-dipole couplings.

As expected, the spin-photon coupling strength $G_{N}$ decreases rapidly with increasing temperature. This dependence can be accounted for by the decrease in spin polarization (cf Fig. \ref{fgr:gvsT}D and \ref{fgr:gvsT}E), showing again the validity of Eq. \ref{eqn:GN}. From these data, and using the topographic and geometrical information on the sample, we estimate an average single spin coupling $G_{1} / 2 \pi \simeq 183$ Hz. This represents an enhancement of nearly two orders of magnitude with respect to the coupling achieved for a $14 \mu$m wide transmission line resonator. Taking into account that the photon energy, thus also $G_{1} \propto \omega_{\rm r}$, it is useful to define the dimensionless coupling $G_{1}/\omega_{\rm r} \simeq 1.3 \times 10^{-7}$. This result compares favourably to the maximum single-spin coupling reported in the literature, of about $450$ Hz for a $7$ GHz resonator (for a ratio $G_{1}/\omega_{\rm r} \simeq 6.4 \times 10^{-8}$),\cite{Probst2017} which was achieved with especially designed lumped-element resonators coupled to impurity spins in the Si substrate on which these devices were fabricated. Yet, it still falls short of the maximum theoretically attainable enhancement, which should scale as $1/w$ and therefore reach a factor $330$ in this case, thus $G_{1} / 2 \pi$ close to $1$ kHz.\cite{Jenkins2013,Jenkins2016} 

In order better understand these results and, in particular, obtain information on how each molecular spin couples to the resonator, we have performed numerical simulations of the spin-photon interaction for this specific situation. The model uses a three-dimensional map of the number of spins extracted from the combined analysis of SEM and AFM images. The region surrounding the nano-constriction is divided into a grid of  cubic cells with lateral dimensions $d$. A $2D$ projection of a map calculated for the smallest $d=3$ nm is shown in Fig. \ref{fgr:AFM}. The coupling is then evaluated as follows. First, the photon energy $\hbar \omega_{\rm r}$ is used to determine the supercurrent flowing through the constriction. This electrical current generates a magnetic field $\vec{b}(\vec{r})$ at each point in space $\vec{r}$, which was calculated using the 3D-MLSI finite-element computer simulation software \cite{Khapaev2002} as in the example shown in Fig. \ref{fgr:circuit}B. For all molecules in a given cell $i$, the magnetic field is assigned its value $\vec{b}_{i}$ at the center of the cell. Then, the coupling of each cell is calculated with the following expression (for details, see \cite{Jenkins2013})

\begin{equation}
G_{i} = g \mu_{\rm B} \sqrt{n_{i}}\vert \langle m_{S}=+1/2 \vert \vec{b_{i}} \vec{S} \vert m_{S}=-1/2 \rangle \vert   
   \label{eqn:Gni}
\end{equation}

\noindent where $n_{i}$ is the effective number of spins in the cell, calculated at the given temperature, and $m_{S} = \pm 1/2$ denote the two eigenstates associated with opposite projections of the radical spins along the external magnetic field $\vec{H}$. The contributions of all cells can then be combined to estimate the collective coupling of the whole sample as follows \cite{Hummer2012}

\begin{equation}
G_{N} = \sqrt{\sum_{i} G_{i}^{2}}
   \label{eqn:Gni2}
\end{equation} 

For the device shown in Fig. \ref{fgr:AFM}, Eq. (\ref{eqn:Gni2}) gives $G_{N} / 2 \pi \simeq 2.0$ MHz at $44$ mK and an average $G_{1} \simeq 200$ Hz, which agree very well with the experimental values $G_{N} = 1.9$ MHz and $G_{1}=183$ Hz and, as shown in Fig. \ref{fgr:gvsT}, accounts well for the transmission experiments. The difference is in fact smaller than the unavoidable errors associated with the number of molecules and with the influence that crystalline defects have on the free-radicals, which are known to turn a fraction of molecules into a diamagnetic state. Simulations can then help in understanding how these average values ensue from the distribution of spin locations and couplings in the deposit. A projection map showing the maximum coupling obtained for cells located at different $xy$ positions (which mainly correspond to those lying closest to the device surface along the vertical axis $z$) is shown in Fig. \ref{fgr:AFM}C. It confirms that the main contribution to $G_{N}$ comes from those spins located in the closest proximity to the constriction. It also shows that individual couplings significantly larger than the average $G_{1}$ can be achieved. Looking at those spins lying closer to the constriction, we find values as high as $0.6-0.8$ kHz, depending on the exact location with respect to the surface and on the orientation of the external magnetic field (see the section on methods and the supplementary information for additional details). 

\subsection{Spin-photon coupling versus transmission line width}

\begin{figure}
\centering
\includegraphics[width=0.5\columnwidth]{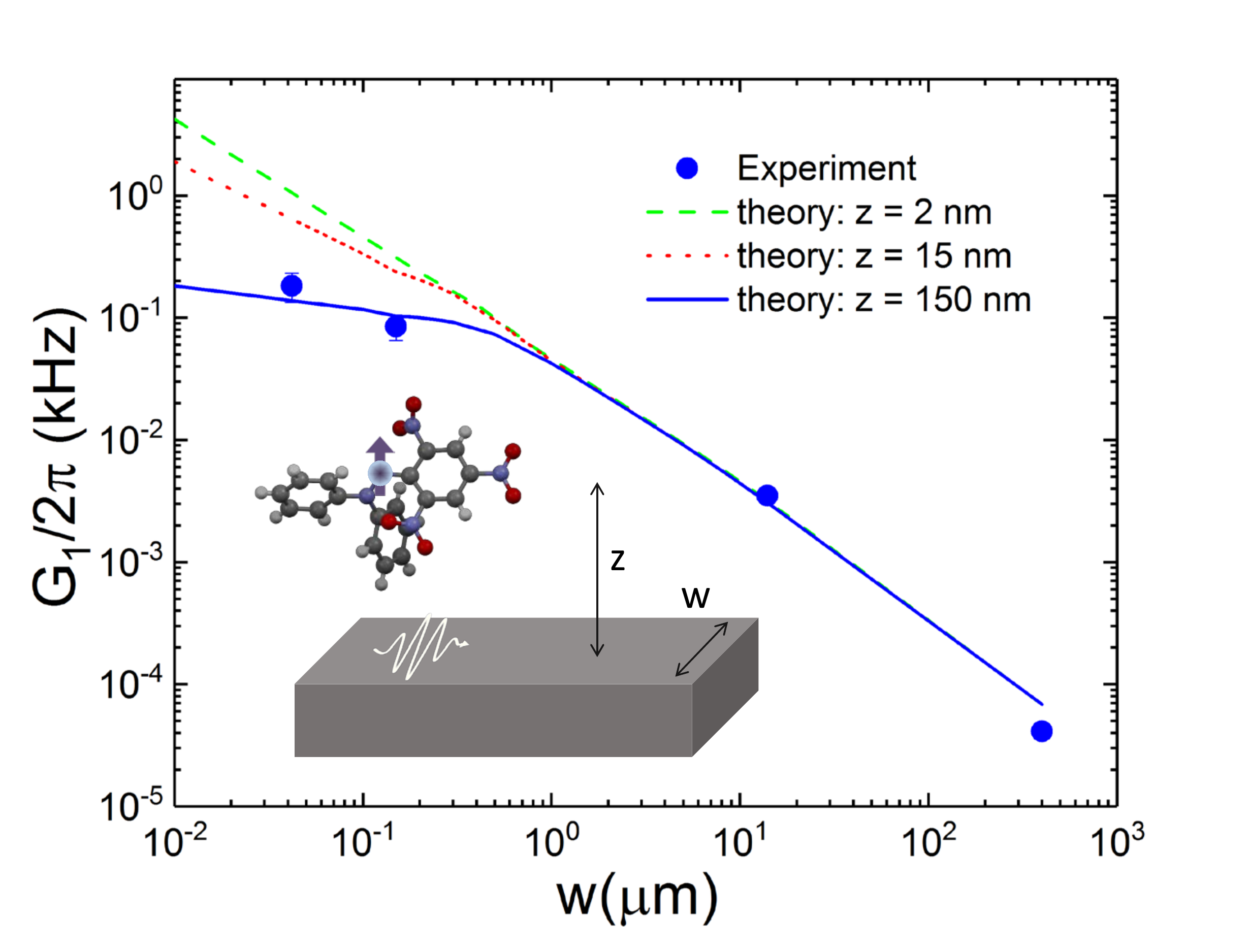}
  \caption{Dependence of the average single spin to single photon coupling on the width of the central transmission line of the resonator, showing the enhancement obtained by reducing the latter down to the region of (tens of) nanometers. The lines are calculations of the coupling of a spin located over the constriction, at three different heights $z$, as illustrated by the figure in the inset.}
  \label{fgr:gvsw}
\end{figure}

Results obtained for different devices enable determining how the single spin to single photon coupling $G_{1}$ depends on the dimensions of the superconducting transmission line. This dependence is shown in Fig. \ref{fgr:gvsw}, which, besides those discussed already, includes also data measured on a resonator with a $400 \mu$m wide line (cf supplementary information). For $w \geqslant 100$ nm, these results show the expected enhancement of $G_{1} \propto 1/w$, from $G_{1} \leqslant 5$ mHz up to nearly $100$ Hz. For narrower lines, however, and as we have already pointed out above, the increase slows down. 

The reason for this effect can be understood on the light of the simulations and arguments described in the previous section (cf Fig. \ref{fgr:AFM}C). Since the coupling enhancement is concentrated close to the constriction, within a region with dimensions comparable to $w$, the sample integration becomes critical. The topography AFM images show that, once dry, DPPH molecules deposited by DPN tend to form nano-aggregates with characteristic dimensions of the order of $200$ nm. Then, even those molecules located in grains that land next to the line can be too far when compared to $w$. This effect can be accounted for by numerical calculations of $G_{1}$. They show that $G_{1}$ strongly depends on the height $z$ of the molecular spins above the transmission line. While, for $z=0$, $G_{1} \propto 1/w$, when $z \neq 0$ the coupling tends to saturate as soon as $w \leqslant z$. Therefore, creating an optimum interface between the molecular deposits and the circuit surface, i.e. making $z$ as close to zero as possible, becomes of utmost importance to attain the maximum spin-photon coupling.

\subsection{Conclusions and future prospects}
The experiments and simulations described in the previous sections validate a relatively straightforward method to enhance the coupling of molecular spins to on-chip superconducting resonators. Using conventional $1.4$ GHz coplanar resonators, we have attained an average single spin coupling of about $183$ Hz for central line widths of $\sim 40$ nm, with estimated maximum couplings of the order of kHz. This approach is, in principle, applicable to any circuit design and to a large variety of samples, provided that they can be delivered from solution and with sufficient spatial accuracy. Introducing superconducting nanobridges into especially designed resonators, which minimize the circuit impedance and therefore maximize the superconducting current at the inductor,\cite{Eichler2017,Probst2017,Sarabi2019} should allow enhancing $G_{1}$ by a further two orders of magnitude, thus reaching $G_{1}/2 \pi$ values close to $0.1$ MHz.

Devices based on these ideas can therefore take electron spin spectroscopy to the single spin limit, opening new avenues for investigating magnetic excitations in individual nanosystems ranging from nanosized particles to molecular nanomagnets, or even exotic topological states, such as vortices and skyrmions.\cite{Martinez2019} The molecular approach used here adds the possibility of serving as a suitable vehicle to deliver diverse samples, as well as to improve their interface with the circuit. A recent work shows that it is possible to carry out a chemical synthesis of thin molecular films onto a superconducting line, thereby achieving a close to optimum coverage with a minimum  molecule to circuit distance.\cite{Urtizberea2020} The combination of these methods with DPN or other nanolithography methods with a high spatial resolution could then lead to major improvements.

Spins hosted in artificial molecules are also very promising candidates to encode qubits and qudits.\cite{Leuenberger2001,Troiani2011,Aromi2012,Moreno-Pineda2018,Gaita2019,Atzori2019} In recent times, spin coherence times of these systems have been optimized by chemical design, up to values well above $10-50 \mu$s,\cite{Ardavan2007,Wedge2012,Bader2014,Shiddiq2016} and even close to ms.\cite{Zadrozny2015}, and the strong coupling to superconducting resonators has been achieved for macroscopic molecular ensembles.\cite{Ghirri2016,Mergenthaler2017,Bonizzoni2017} Our results show that attaining this regime for single molecules is within reach. This coherent coupling would allow the use of superconducting circuits to control, read-out and connect spin qubits located in different molecules, thus paving the way to a hybrid architecture with a huge potential for large scale quantum computation and simulation.\cite{Jenkins2016}     

\section{Methods}

\subsection{Device fabrication and characterization}
Superconducting resonators are fabricated on $500 \mu$m thick C-plane sapphire wafers. A $150$ nm thick niobium layer was deposited
by radio frequency sputtering and then patterned
by photolithography and reactive ion etching. The circuits consist of a large $400 \mu$m central line separated from two ground planes by two $200 \mu$m gaps that narrow down to around $14 \mu$m and $7\mu$m, respectively, after going through gap capacitors. The devices were tuned to show an effective impedance $Z_{0} = 50 \Omega$, a resonance frequency $\omega_{\rm r}/2 \pi \simeq 1.4$ GHz and maximum quality factors $Q \simeq 1-2\times 10^{5}$. Nanoscale constrictions were made at the midpoint of the central line by etching it with a focused beam of Ga$^{+}$ ions, using a commercial dual beam system. The ionic current was kept below $20$ pA to maximize the resolution in the fabrication process and to minimize the Nb thickness, of order $10$ to $15$ nm, that is implanted with Ga. In order to avoid the build-up, and eventual discharge, of electrostatic charges during the process, the central line was connected to one of the ground planes by a few nm wide Pt bridge, fabricated by focused ion beam deposition. Once the nanoconstriction was fabricated, this bridge was removed using the same ion beam. Images of $158$ nm and $42$ nm wide constrictions, obtained in situ by SEM, are shown in Fig. \ref{fgr:circuit} and in the supplementary information.

Prior to the deposition of the molecular samples, the microwave propagation trough all devices was characterized, as described below, as a function of magnetic field, temperature and input power. Illustrative results are shown as part of the supplementary information. These experiments show that the constrictions do not introduce any drastic changes to $\omega_{\rm r}$ and $Q$, and that they can work under magnetic fields up to $1.4$ T, provided that they are applied parallel to the plane of the chip.

\subsection{Molecular pattering onto superconducting resonators and characterization of the deposits}
The molecular ink used in all depositions was prepared by dissolving $10-20$ mg/ml of the free radical 2,2-diphenyl-1-picrylhydrazyl (DPPH; Sigma-Aldrich) in N,N-Dimethylformamide (DMF; Sigma-Aldrich) with about $5-10$ \% glycerol in volume (Panreac, $\geq 99.5$ \%, ACS grade). DMF preserves the chemical properties of DPPH and does not turn the radical into its diamagnetic form. Glycerol is used as additive to increase the viscosity and slow down the evaporation of the ink so that the solution does not dry before the creation of the pattern.\cite{Bellido2010} This is especially important in the deposition by DPN, when the drop has to be transferred to the device by the AFM tip. The device was previously cleaned with isopropanol and acetone, dried by blowing nitrogen gas and underwent plasma oxygen treatment before deposition. DPN deposition was performed using a DPN5000 based-AFM system (NanoInk, Inc., USA). For this, $1\mu$l of the ink solution was left into a reservoir of a microfluidic Inkwell delivery chip-based system (Acs-technologies LLC, USA). Afterwards, a DPN silicon nitride DPN single A-S2 probe (Acs-technologies LLC, USA) was dipped and coated several times with the ink at the inkwell micro-channel. DPN patterning was optimized at $25^{\circ}$C and $55$ \% relative humidity. The quality and size of the free radical deposits were assayed previously on marked SiO$_{2}$ substrates and small pieces of muscovite mica (see SI for illustrative images of the results). The deposition conditions were optimized to produce deposits ranging from $1$ to $60 \mu$m in diameter, such as the one shown in Fig. S8. Then, different drops with diameters ranging between $5-60 \mu$m and heights from $50$ to $400$ nm were patterned on the central line of resonators, both without constriction and with nanoconstrictions. The characterization of the deposits and the microwave transmission experiments were performed once the deposits had dried. Free radical deposits were also made far from the sensing areas and using different solvents and additives and characterized as controls. 

The patterns were analyzed by optical microscopy, AFM and SEM imaging. AFM images were taken with a MultiMode $8$ AFM system (Bruker) and a Cervantes SPM FullMode (Nanotec) at scan rates of $0.1-2.0$ Hz. The samples were characterized using Peak Force TappingTM and TappingTM modes with $7-125$ kHz triangular silicon nitride microlevers (SNL; Bruker Probes) having ultrasharp $2$ nm end tip radii, and with antimony (n) doped silicon $320$ kHz ultraresonant microlever (TESP; Bruker Probes) having $8$ nm end tip radii. The images were analyzed using the WSxM \cite{Horcas2007} and Gwyddion \cite{Necas2011} softwares for SPM image processing. SEM images were performed with an INSPECT-F50 (FEI). 

\subsection{Microwave transmission experiments}

The microwave transmission measurements were done using a programmable network analyzer. Experiments performed at $4.2$ K were aimed at measuring a large number of deposits of varying dose. In these experiments, the devices were mounted on a home-made probe and submerged in a liquid helium cryostat. The input power was attenuated by $-52$dB, to avoid the possibility of exceeding the critical current at the nanoconstriction, and the output power was measured directly. The number of photons in the cavity was estimated to be of order $10^9$. The external magnetic field was applied with a commercial $9$T$\times 1$T$\times 1$T superconducting vector magnet, which allows rotating $\vec{H}$ in situ, in order to align it along the central line of the device ($y$ axis in Figs. \ref{fgr:circuit} and \ref{fgr:AFM}) with an accuracy better than $0.1$ degrees. For experiments at lower temperatures, transmission measurements were carried out in an adiabatic demagnetization refrigerator, having a base temperature of about $45$ mK. Magnetic fields up to $60$ mT were applied using a home made superconducting vector magnet. In this case, experimental constraints imposed a field orientation along the $x$ axis, thus perpendicular to the central line within the plane of the device. As it is shown in the supplementary information, this geometry leads to a substantial decrease in the spin-photon coupling with respect to the maximum achievable. The input signal was attenuated, down to a level of about $5\times10^{5}$ photons, and then amplified at $4.2$ K and at room temperature before being detected.

\subsection{Numerical simulations}
Our estimate of the spin-photon coupling relies on the calculation of the spatial distribution of $\vec b (\vec r)$, i.e., the rms magnetic field generated by the rms supercurrent $i_{\rm sc} = \omega_{\rm r} \sqrt{ \pi \hbar / 4 Z_0} = 11.3$ nA circulating through the central line of the resonator, with $Z_0 \sim 50$ $\Omega$ its impedance.\citep{Jenkins2013} The spatial distribution of $i_{\rm sc}(\vec r)$ is computed using a finite-element based package\citep{Khapaev2002} that solves the London equations for the given geometry of the superconducting wire, having London penetration depth $\lambda_{\rm L}=90$ nm, and assuming that the supercurrent distributes through $11$ equidistant 2D sheets in the \textit{xy} plane. $i_{\rm sc}(\vec r)$ is then used to evaluate the rms magnetic field $\vec b_i$ (with $i=1,2...N_{\rm cell}$) at the center of $N_{\rm cell}$ cubic cells with lateral dimension $d$, each containing $n_i$ molecules. The grid size was varied from $0$ up to $2 \mu$m and the cell size $d$ from $3$ nm up to $100$ nm. Illustrative results are shown as part of the supplementary information. These simulations show that spins located further than about $500$ nm from the constriction give a close to negligible contribution. The mode volume of a $42$ nm wide constriction is then $\simeq 1 \mu$m $\times 1 \mu$m $\times 10 \mu$m $=10 \mu$m$^{3} = 0.01$pico-l, or about $1.4 \times 10^{-14} \lambda^{3}$, with $\lambda \simeq 88$ mm being the resonance wave length. Also, that a $3$ nm cell size is necessary to properly account for the coupling to spins that lie very close to the nanoconstriction, which give the maximum contribution. The maximum coupling $G_{1}$ depends also on the orientation of the external magnetic field, which determines which components of the photon magnetic field couple to the spins. It is maximum for $\vec{H}$ parallel to the central line ($y$ axis). Further details can be found in the supplementary information file.

\section{About this version and final publication}

This document is the unedited Author's version of a Submitted Work that was subsequently accepted for publication in ACS Nano, copyright © American Chemical Society after peer review. To access the final edited and published work see http://dx.doi.org/10.1021/acsnano.0c03167.

\begin{acknowledgement}

The authors acknowledge funding from the EU (COST Action $15128$ MOLSPIN, QUANTERA SUMO and MICROSENSE projects, FET-OPEN grant $862893$ FATMOLS), the Spanish MICINN (grants RTI2018-096075-B-C21, PCI2018-093116, MAT2017-89993-R, MAT2017-88358-C3-1-R, EUR2019-103823), the Gobierno de Arag\'on (grants E09-17R Q-MAD, E35-20R, BE and LMP55-18, FANDEPAM) and the BBVA foundation (Leonardo grants 2018 and 2019).

\end{acknowledgement}

\begin{suppinfo}

The following file is available free of charge.
\begin{itemize}
  \item GimenoACSnanoSI.pdf: Includes details about the fabrication and test of coplanar superconducting resonators and of superconducting nanoconstrictions, the deposition of free-radical spin ensembles by DPN, their characterization by means of SEM and AFM, results of additional microwave transmission experiments, the calculation of the number of molecules effectively coupled to each device and the results of simulations of the spin-photon coupling. 
\end{itemize}

\end{suppinfo}


\bibliography{bibliography_Gimeno_ACSnano}

\providecommand{\latin}[1]{#1}
\makeatletter
\providecommand{\doi}
  {\begingroup\let\do\@makeother\dospecials
  \catcode`\{=1 \catcode`\}=2 \doi@aux}
\providecommand{\doi@aux}[1]{\endgroup\texttt{#1}}
\makeatother
\providecommand*\mcitethebibliography{\thebibliography}
\csname @ifundefined\endcsname{endmcitethebibliography}
  {\let\endmcitethebibliography\endthebibliography}{}
\begin{mcitethebibliography}{59}
\providecommand*\natexlab[1]{#1}
\providecommand*\mciteSetBstSublistMode[1]{}
\providecommand*\mciteSetBstMaxWidthForm[2]{}
\providecommand*\mciteBstWouldAddEndPuncttrue
  {\def\EndOfBibitem{\unskip.}}
\providecommand*\mciteBstWouldAddEndPunctfalse
  {\let\EndOfBibitem\relax}
\providecommand*\mciteSetBstMidEndSepPunct[3]{}
\providecommand*\mciteSetBstSublistLabelBeginEnd[3]{}
\providecommand*\EndOfBibitem{}
\mciteSetBstSublistMode{f}
\mciteSetBstMaxWidthForm{subitem}{(\alph{mcitesubitemcount})}
\mciteSetBstSublistLabelBeginEnd
  {\mcitemaxwidthsubitemform\space}
  {\relax}
  {\relax}

\bibitem[Wallraff \latin{et~al.}(2004)Wallraff, Schuster, Blais, Frunzio,
  Huang, Majer, Kumar, Girvin, and Schoelkopf]{Wallraff2004}
Wallraff,~A.; Schuster,~D.~I.; Blais,~A.; Frunzio,~L.; Huang,~R.-S.; Majer,~J.;
  Kumar,~S.; Girvin,~S.~M.; Schoelkopf,~R.~J. {Strong Coupling of a Single
  Photon to a Superconducting Qubit Using Circuit Quantum Electrodynamics}.
  \emph{Nature} \textbf{2004}, \emph{431}, 162--167\relax
\mciteBstWouldAddEndPuncttrue
\mciteSetBstMidEndSepPunct{\mcitedefaultmidpunct}
{\mcitedefaultendpunct}{\mcitedefaultseppunct}\relax
\EndOfBibitem
\bibitem[G{\"o}ppl \latin{et~al.}(2008)G{\"o}ppl, Fragner, Baur, Bianchetti,
  Filipp, Fink, Leek, Puebla, Steffen, and Wallraff]{Goppl2008}
G{\"o}ppl,~M.; Fragner,~A.; Baur,~M.; Bianchetti,~R.; Filipp,~S.; Fink,~J.~M.;
  Leek,~P.~J.; Puebla,~G.; Steffen,~L.; Wallraff,~A. {Coplanar Waveguide
  Resonators for Circuit Quantum Electrodynamics}. \emph{J. Appl. Phys.}
  \textbf{2008}, \emph{104}, 11394\relax
\mciteBstWouldAddEndPuncttrue
\mciteSetBstMidEndSepPunct{\mcitedefaultmidpunct}
{\mcitedefaultendpunct}{\mcitedefaultseppunct}\relax
\EndOfBibitem
\bibitem[Bienfait \latin{et~al.}(2016)Bienfait, Pla, Kubo, Stern, Lo, Weis,
  Schenkel, Thewalt, Vion, Esteve, Julsgaard, M\o{}lmer, Morton, and
  Bertet]{Bienfait2016}
Bienfait,~A.; Pla,~A.~A.; Kubo,~Y.; Stern,~X.,~M.~Zhou; Lo,~C.~C.; Weis,~C.~D.;
  Schenkel,~T.; Thewalt,~M. L.~W.; Vion,~D.; Esteve,~D.; Julsgaard,~B.;
  M\o{}lmer,~K.; Morton,~J. J.~L.; Bertet,~P. {Reaching the Quantum Limit of
  Sensitivity in Electron Spin Resonance}. \emph{Nat. Nanotechnol.}
  \textbf{2016}, \emph{11}, 253--257\relax
\mciteBstWouldAddEndPuncttrue
\mciteSetBstMidEndSepPunct{\mcitedefaultmidpunct}
{\mcitedefaultendpunct}{\mcitedefaultseppunct}\relax
\EndOfBibitem
\bibitem[Eichler \latin{et~al.}(2017)Eichler, Sigillito, Lyon, and
  Petta]{Eichler2017}
Eichler,~C.; Sigillito,~A.~J.; Lyon,~S.~A.; Petta,~J.~R. {Electron Spin
  Resonance at the Level of $10^{4}$ Spins Using Low Impedance Superconducting
  Resonators}. \emph{Phys. Rev. Lett.} \textbf{2017}, \emph{118}, 037701\relax
\mciteBstWouldAddEndPuncttrue
\mciteSetBstMidEndSepPunct{\mcitedefaultmidpunct}
{\mcitedefaultendpunct}{\mcitedefaultseppunct}\relax
\EndOfBibitem
\bibitem[Probst \latin{et~al.}(2017)Probst, Bienfait, Campagne-Ibarcq, Pla,
  Albanese, Da~Silva~Barbosa, Schenkel, Vion, Esteve, M\o{}lmer, Morton,
  Heeres, and Bertet]{Probst2017}
Probst,~S.; Bienfait,~A.; Campagne-Ibarcq,~P.; Pla,~J.~J.; Albanese,~B.;
  Da~Silva~Barbosa,~J.~F.; Schenkel,~T.; Vion,~D.; Esteve,~D.; M\o{}lmer,~K.;
  Morton,~J. J.~L.; Heeres,~R.; Bertet,~P. {Inductive-Detection Electron-Spin
  Resonance spectroscopy with 65 spins/ Hz sensitivity}. \emph{Appl. Phys.
  Lett.} \textbf{2017}, \emph{111}, 202604\relax
\mciteBstWouldAddEndPuncttrue
\mciteSetBstMidEndSepPunct{\mcitedefaultmidpunct}
{\mcitedefaultendpunct}{\mcitedefaultseppunct}\relax
\EndOfBibitem
\bibitem[Sarabi \latin{et~al.}(2019)Sarabi, Huang, and Zimmerman]{Sarabi2019}
Sarabi,~B.; Huang,~P.; Zimmerman,~N.~M. {Possible Hundredfold Enhancement in
  the Direct Magnetic Coupling of a Single-Atom Electron Spin to a Circuit
  Resonator}. \emph{Phys. Rev. Appl.} \textbf{2019}, \emph{11}, 014001\relax
\mciteBstWouldAddEndPuncttrue
\mciteSetBstMidEndSepPunct{\mcitedefaultmidpunct}
{\mcitedefaultendpunct}{\mcitedefaultseppunct}\relax
\EndOfBibitem
\bibitem[Blais \latin{et~al.}(2004)Blais, Huang, Wallraff, Girvin, and
  Schoelkopf]{Blais2004}
Blais,~A.; Huang,~R.-S.; Wallraff,~A.; Girvin,~S.~M.; Schoelkopf,~R.~J. {Cavity
  Quantum Electrodynamics for Superconducting Electrical Circuits: An
  Architecture for Quantum Computation}. \emph{Phys. Rev. A} \textbf{2004},
  \emph{69}, 062320\relax
\mciteBstWouldAddEndPuncttrue
\mciteSetBstMidEndSepPunct{\mcitedefaultmidpunct}
{\mcitedefaultendpunct}{\mcitedefaultseppunct}\relax
\EndOfBibitem
\bibitem[Majer \latin{et~al.}(2007)Majer, Chow, Gambetta, Koch, Johnson,
  Schreier, Frunzio, Schuster, Houck, Wallraff, Blais, Devoret, Girvin, and
  Schoelkopf]{Majer2007}
Majer,~J.; Chow,~J.~M.; Gambetta,~J.~M.; Koch,~J.; Johnson,~B.~R.;
  Schreier,~J.~A.; Frunzio,~L.; Schuster,~D.~I.; Houck,~A.~A.; Wallraff,~A.;
  Blais,~A.; Devoret,~M.~H.; Girvin,~S.~M.; Schoelkopf,~R.~J. Coupling
  Superconducting Qubits {\em via} a Cavity Bus. \emph{Nature} \textbf{2007},
  \emph{449}, 443--447\relax
\mciteBstWouldAddEndPuncttrue
\mciteSetBstMidEndSepPunct{\mcitedefaultmidpunct}
{\mcitedefaultendpunct}{\mcitedefaultseppunct}\relax
\EndOfBibitem
\bibitem[Schoelkopf and Girvin(2008)Schoelkopf, and Girvin]{Schoelkopf2008}
Schoelkopf,~R.~J.; Girvin,~S.~M. Wiring Up Quantum Systems. \emph{Nature}
  \textbf{2008}, \emph{451}, 664--669\relax
\mciteBstWouldAddEndPuncttrue
\mciteSetBstMidEndSepPunct{\mcitedefaultmidpunct}
{\mcitedefaultendpunct}{\mcitedefaultseppunct}\relax
\EndOfBibitem
\bibitem[Imamo\v{g}lu(2009)]{Imamoglu2009}
Imamo\v{g}lu,~A. {Cavity QED Based on Collective Magnetic Dipole Coupling: Spin
  Ensembles as Hybrid Two-Level Systems}. \emph{Phys. Rev. Lett.}
  \textbf{2009}, \emph{102}, 083602\relax
\mciteBstWouldAddEndPuncttrue
\mciteSetBstMidEndSepPunct{\mcitedefaultmidpunct}
{\mcitedefaultendpunct}{\mcitedefaultseppunct}\relax
\EndOfBibitem
\bibitem[Wesenberg \latin{et~al.}(2009)Wesenberg, Ardavan, Briggs, Morton,
  Schoelkopf, Schuster, and M\o{}lmer]{Wesenberg2009}
Wesenberg,~J.~H.; Ardavan,~A.; Briggs,~G. A.~D.; Morton,~J. J.~L.;
  Schoelkopf,~R.~J.; Schuster,~D.~I.; M\o{}lmer,~K. {Quantum Computing with an
  Electron Spin Ensemble}. \emph{Phys. Rev. Lett.} \textbf{2009}, \emph{103},
  070502\relax
\mciteBstWouldAddEndPuncttrue
\mciteSetBstMidEndSepPunct{\mcitedefaultmidpunct}
{\mcitedefaultendpunct}{\mcitedefaultseppunct}\relax
\EndOfBibitem
\bibitem[Schuster \latin{et~al.}(2010)Schuster, Sears, Ginossar, DiCarlo,
  Frunzio, Morton, Wu, Briggs, Buckley, Awschalom, and
  Schoelkopf]{Schuster2010}
Schuster,~D.~I.; Sears,~A.~P.; Ginossar,~E.; DiCarlo,~L.; Frunzio,~L.;
  Morton,~J. J.~L.; Wu,~H.; Briggs,~G. A.~D.; Buckley,~B.~B.; Awschalom,~D.~D.;
  Schoelkopf,~R.~J. {High-Cooperativity Coupling of Electron-Spin Ensembles to
  Superconducting Cavities}. \emph{Phys. Rev. Lett.} \textbf{2010}, \emph{105},
  140501\relax
\mciteBstWouldAddEndPuncttrue
\mciteSetBstMidEndSepPunct{\mcitedefaultmidpunct}
{\mcitedefaultendpunct}{\mcitedefaultseppunct}\relax
\EndOfBibitem
\bibitem[Kubo \latin{et~al.}(2010)Kubo, Ong, Bertet, Vion, Jacques, Zheng,
  Dr\'eau, Roch, Auffeves, Jelezko, Wrachtrup, Barthe, Bergonzo, and
  Esteve]{Kubo2010}
Kubo,~Y.; Ong,~F.~R.; Bertet,~P.; Vion,~D.; Jacques,~V.; Zheng,~D.;
  Dr\'eau,~A.; Roch,~J.-F.; Auffeves,~A.; Jelezko,~F.; Wrachtrup,~J.;
  Barthe,~M.~F.; Bergonzo,~P.; Esteve,~D. {Strong Coupling of a Spin Ensemble
  to a Superconducting Resonator}. \emph{Phys. Rev. Lett.} \textbf{2010},
  \emph{105}, 140502\relax
\mciteBstWouldAddEndPuncttrue
\mciteSetBstMidEndSepPunct{\mcitedefaultmidpunct}
{\mcitedefaultendpunct}{\mcitedefaultseppunct}\relax
\EndOfBibitem
\bibitem[Wu \latin{et~al.}(2010)Wu, George, Wesenberg, M\o{}lmer, Schuster,
  Schoelkopf, Itoh, Ardavan, Morton, and Briggs]{Wu2010}
Wu,~H.; George,~R.~E.; Wesenberg,~J.~H.; M\o{}lmer,~K.; Schuster,~D.~I.;
  Schoelkopf,~R.~J.; Itoh,~K.~M.; Ardavan,~A.; Morton,~J. J.~L.; Briggs,~G.
  A.~D. {Storage of Multiple Coherent Microwave Excitations in an Electron Spin
  Ensemble}. \emph{Phys. Rev. Lett.} \textbf{2010}, \emph{105}, 140503\relax
\mciteBstWouldAddEndPuncttrue
\mciteSetBstMidEndSepPunct{\mcitedefaultmidpunct}
{\mcitedefaultendpunct}{\mcitedefaultseppunct}\relax
\EndOfBibitem
\bibitem[Chiorescu \latin{et~al.}(2010)Chiorescu, Groll, Bertaina, Mori, and
  Miyashita]{Chiorescu2010}
Chiorescu,~I.; Groll,~N.; Bertaina,~S.; Mori,~T.; Miyashita,~S. {Magnetic
  Strong Coupling in a Spin-Photon System and Transition to Classical Regime}.
  \emph{Phys. Rev. B} \textbf{2010}, \emph{82}, 024413\relax
\mciteBstWouldAddEndPuncttrue
\mciteSetBstMidEndSepPunct{\mcitedefaultmidpunct}
{\mcitedefaultendpunct}{\mcitedefaultseppunct}\relax
\EndOfBibitem
\bibitem[Jenkins \latin{et~al.}(2016)Jenkins, Zueco, Roubeau, Arom\'i, Majer,
  and Luis]{Jenkins2016}
Jenkins,~M.~D.; Zueco,~D.; Roubeau,~O.; Arom\'i,~G.; Majer,~J.; Luis,~F. {A
  Scalable Architecture for Quantum Computation with Molecular Nanomagnets}.
  \emph{Dalton Trans.} \textbf{2016}, \emph{45}, 16682--16693\relax
\mciteBstWouldAddEndPuncttrue
\mciteSetBstMidEndSepPunct{\mcitedefaultmidpunct}
{\mcitedefaultendpunct}{\mcitedefaultseppunct}\relax
\EndOfBibitem
\bibitem[Narkowicz \latin{et~al.}(2005)Narkowicz, Suter, and
  Stonies]{Narkowicz2005}
Narkowicz,~R.; Suter,~D.; Stonies,~R. {Planar Microresonators for EPR
  Experiments}. \emph{J. Magn. Reson.} \textbf{2005}, \emph{175},
  275--284\relax
\mciteBstWouldAddEndPuncttrue
\mciteSetBstMidEndSepPunct{\mcitedefaultmidpunct}
{\mcitedefaultendpunct}{\mcitedefaultseppunct}\relax
\EndOfBibitem
\bibitem[Narkowicz \latin{et~al.}(2008)Narkowicz, Suter, and
  Niemeyer]{Narkowicz2008}
Narkowicz,~R.; Suter,~D.; Niemeyer,~I. {Scaling of Sensitivity and Efficiency
  in Planar Microresonators for Electron Spin Resonance}. \emph{Rev. Sci.
  Instrum.} \textbf{2008}, \emph{79}, 084702\relax
\mciteBstWouldAddEndPuncttrue
\mciteSetBstMidEndSepPunct{\mcitedefaultmidpunct}
{\mcitedefaultendpunct}{\mcitedefaultseppunct}\relax
\EndOfBibitem
\bibitem[Banholzer \latin{et~al.}(2011)Banholzer, Narkowicz, Hassel,
  Meckenstock, Stienen, Posth, Suter, Farle, and Lindner]{Banholzer2011}
Banholzer,~A.; Narkowicz,~R.; Hassel,~C.; Meckenstock,~R.; Stienen,~S.;
  Posth,~O.; Suter,~D.; Farle,~M.; Lindner,~J. {Visualization of Spin Dynamics
  in Single Nanosized Magnetic Elements}. \emph{Nanotechnology} \textbf{2011},
  \emph{22}, 295713\relax
\mciteBstWouldAddEndPuncttrue
\mciteSetBstMidEndSepPunct{\mcitedefaultmidpunct}
{\mcitedefaultendpunct}{\mcitedefaultseppunct}\relax
\EndOfBibitem
\bibitem[Jenkins \latin{et~al.}(2013)Jenkins, H{\"u}mmer, Mart\'inez-P\'erez,
  Garc\'ia-Ripoll, Zueco, and Luis]{Jenkins2013}
Jenkins,~M.~D.; H{\"u}mmer,~T.; Mart\'inez-P\'erez,~M.~J.; Garc\'ia-Ripoll,~J.;
  Zueco,~D.; Luis,~F. {Coupling Single-Molecule Magnets to Quantum Circuits}.
  \emph{New J. Phys.} \textbf{2013}, \emph{15}, 095007\relax
\mciteBstWouldAddEndPuncttrue
\mciteSetBstMidEndSepPunct{\mcitedefaultmidpunct}
{\mcitedefaultendpunct}{\mcitedefaultseppunct}\relax
\EndOfBibitem
\bibitem[Jenkins \latin{et~al.}(2014)Jenkins, Naether, Ciria, Ses\'e, Atkinson,
  S\'anchez-Azqueta, Barco, Majer, Zueco, and Luis]{Jenkins2014}
Jenkins,~M.~D.; Naether,~U.; Ciria,~M.; Ses\'e,~J.; Atkinson,~J.;
  S\'anchez-Azqueta,~C.; Barco,~E.~d.; Majer,~J.; Zueco,~D.; Luis,~F.
  {Nanoscale Constrictions in Superconducting Coplanar Waveguide Resonators}.
  \emph{Appl. Phys. Lett.} \textbf{2014}, \emph{105}, 162601\relax
\mciteBstWouldAddEndPuncttrue
\mciteSetBstMidEndSepPunct{\mcitedefaultmidpunct}
{\mcitedefaultendpunct}{\mcitedefaultseppunct}\relax
\EndOfBibitem
\bibitem[Haikka \latin{et~al.}(2017)Haikka, Kubo, Bienfait, Bertet, and
  M\o{}lmer]{Haikka2017}
Haikka,~P.; Kubo,~Y.; Bienfait,~A.; Bertet,~P.; M\o{}lmer,~K. {Proposal for
  Detecting a Single Electron Spin in a Microwave Resonator}. \emph{Phys. Rev.
  A} \textbf{2017}, \emph{95}, 022306\relax
\mciteBstWouldAddEndPuncttrue
\mciteSetBstMidEndSepPunct{\mcitedefaultmidpunct}
{\mcitedefaultendpunct}{\mcitedefaultseppunct}\relax
\EndOfBibitem
\bibitem[Mannini \latin{et~al.}(2009)Mannini, Pineider, Sainctavit, Danieli,
  Otero, Sciancalepore, Talarico, Arrio, Cornia, Gatteschi, and
  Sessoli]{Mannini2009}
Mannini,~M.; Pineider,~F.; Sainctavit,~P.; Danieli,~C.; Otero,~E.;
  Sciancalepore,~C.; Talarico,~A.~M.; Arrio,~M.-A.; Cornia,~A.; Gatteschi,~D.;
  Sessoli,~R. {Magnetic Memory of a Single-Molecule Quantum Magnet Wired to a
  Gold Surface}. \emph{Nat. Mater.} \textbf{2009}, \emph{8}, 194--197\relax
\mciteBstWouldAddEndPuncttrue
\mciteSetBstMidEndSepPunct{\mcitedefaultmidpunct}
{\mcitedefaultendpunct}{\mcitedefaultseppunct}\relax
\EndOfBibitem
\bibitem[Mannini \latin{et~al.}(2010)Mannini, Pineider, Danieli, Totti, Sorace,
  Sainctavit, Arrio, Otero, Joly, Cezar, Cornia, and Sessoli]{Mannini2010}
Mannini,~M.; Pineider,~F.; Danieli,~C.; Totti,~F.; Sorace,~L.; Sainctavit,~P.;
  Arrio,~M.-A.; Otero,~E.; Joly,~L.; Cezar,~J.~C.; Cornia,~A.; Sessoli,~R.
  {Quantum Tunnelling of the Magnetization in a Monolayer of Oriented
  Single-Molecule Magnets}. \emph{Nature} \textbf{2010}, \emph{468},
  417--421\relax
\mciteBstWouldAddEndPuncttrue
\mciteSetBstMidEndSepPunct{\mcitedefaultmidpunct}
{\mcitedefaultendpunct}{\mcitedefaultseppunct}\relax
\EndOfBibitem
\bibitem[Domingo \latin{et~al.}(2012)Domingo, Bellido, and
  Ruiz-Molina]{Domingo2012}
Domingo,~N.; Bellido,~E.; Ruiz-Molina,~D. Advances on structuring{,}
  Integration and Magnetic Characterization of Molecular Nanomagnets on
  Surfaces and Devices. \emph{Chem. Soc. Rev.} \textbf{2012}, \emph{41},
  258--302\relax
\mciteBstWouldAddEndPuncttrue
\mciteSetBstMidEndSepPunct{\mcitedefaultmidpunct}
{\mcitedefaultendpunct}{\mcitedefaultseppunct}\relax
\EndOfBibitem
\bibitem[Thiele \latin{et~al.}(2014)Thiele, Balestro, Ballou, Klyatskaya,
  Ruben, and Wernsdorfer]{Thiele2014}
Thiele,~S.; Balestro,~F.; Ballou,~R.; Klyatskaya,~S.; Ruben,~M.;
  Wernsdorfer,~W. {Electrically Driven Nuclear Spin Resonance in
  Single-Molecule Magnets}. \emph{Science} \textbf{2014}, \emph{344},
  1135--1138\relax
\mciteBstWouldAddEndPuncttrue
\mciteSetBstMidEndSepPunct{\mcitedefaultmidpunct}
{\mcitedefaultendpunct}{\mcitedefaultseppunct}\relax
\EndOfBibitem
\bibitem[Malavolti \latin{et~al.}(2018)Malavolti, Briganti, H{\"a}nze, Serrano,
  Cimatti, McMurtrie, Otero, Ohresser, Totti, Mannini, Sessoli, and
  Loth]{Malavolti2018}
Malavolti,~L.; Briganti,~M.; H{\"a}nze,~M.; Serrano,~G.; Cimatti,~I.;
  McMurtrie,~G.; Otero,~E.; Ohresser,~P.; Totti,~F.; Mannini,~M.; Sessoli,~R.;
  Loth,~S. {Tunable Spin--Superconductor Coupling of Spin 1/2 Vanadyl
  Phthalocyanine Molecules}. \emph{Nano Lett.} \textbf{2018}, \emph{18},
  7955\relax
\mciteBstWouldAddEndPuncttrue
\mciteSetBstMidEndSepPunct{\mcitedefaultmidpunct}
{\mcitedefaultendpunct}{\mcitedefaultseppunct}\relax
\EndOfBibitem
\bibitem[Urtizberea \latin{et~al.}(2020)Urtizberea, Natividad, Alonso,
  P\'erez-Mart\'inez, Andr\'es, Gasc\'on, Gimeno, Luis, and
  Roubeau]{Urtizberea2020}
Urtizberea,~A.; Natividad,~E.; Alonso,~P.~J.; P\'erez-Mart\'inez,~L.;
  Andr\'es,~M.~A.; Gasc\'on,~I.; Gimeno,~I.; Luis,~F.; Roubeau,~O. {Vanadyl
  Spin Qubit 2D Arrays and their Integration on Superconducting Resonators}.
  \emph{Mater. Horiz.} \textbf{2020}, \emph{7}, 885--897\relax
\mciteBstWouldAddEndPuncttrue
\mciteSetBstMidEndSepPunct{\mcitedefaultmidpunct}
{\mcitedefaultendpunct}{\mcitedefaultseppunct}\relax
\EndOfBibitem
\bibitem[Leuenberger and Loss(2001)Leuenberger, and Loss]{Leuenberger2001}
Leuenberger,~M.; Loss,~D. {Quantum Computing in Molecular Magnets}.
  \emph{Nature} \textbf{2001}, \emph{410}, 789--793\relax
\mciteBstWouldAddEndPuncttrue
\mciteSetBstMidEndSepPunct{\mcitedefaultmidpunct}
{\mcitedefaultendpunct}{\mcitedefaultseppunct}\relax
\EndOfBibitem
\bibitem[Troiani and Affronte(2011)Troiani, and Affronte]{Troiani2011}
Troiani,~F.; Affronte,~M. {Molecular Spins for Quantum Information
  technologies}. \emph{Chem. Soc. Rev.} \textbf{2011}, \emph{40},
  3119--3129\relax
\mciteBstWouldAddEndPuncttrue
\mciteSetBstMidEndSepPunct{\mcitedefaultmidpunct}
{\mcitedefaultendpunct}{\mcitedefaultseppunct}\relax
\EndOfBibitem
\bibitem[Arom\'i \latin{et~al.}(2012)Arom\'i, Aguil\`a, Gamez, Luis, and
  Roubeau]{Aromi2012}
Arom\'i,~G.; Aguil\`a,~D.; Gamez,~P.; Luis,~F.; Roubeau,~O. {Design of Magnetic
  Coordination Complexes for Quantum Computing}. \emph{Chem. Soc. Rev.}
  \textbf{2012}, \emph{41}, 537--546\relax
\mciteBstWouldAddEndPuncttrue
\mciteSetBstMidEndSepPunct{\mcitedefaultmidpunct}
{\mcitedefaultendpunct}{\mcitedefaultseppunct}\relax
\EndOfBibitem
\bibitem[Moreno-Pineda \latin{et~al.}(2018)Moreno-Pineda, Godfrin, Balestro,
  Wernsdorfer, and Ruben]{Moreno-Pineda2018}
Moreno-Pineda,~E.; Godfrin,~C.; Balestro,~F.; Wernsdorfer,~W.; Ruben,~M.
  {Molecular Spin qudits for Quantum Algorithms}. \emph{Chem. Soc. Rev.}
  \textbf{2018}, \emph{47}, 501--513\relax
\mciteBstWouldAddEndPuncttrue
\mciteSetBstMidEndSepPunct{\mcitedefaultmidpunct}
{\mcitedefaultendpunct}{\mcitedefaultseppunct}\relax
\EndOfBibitem
\bibitem[Gaita-Ari{\~n}o \latin{et~al.}(2019)Gaita-Ari{\~n}o, Luis, Hill, and
  Coronado]{Gaita2019}
Gaita-Ari{\~n}o,~A.; Luis,~F.; Hill,~S.; Coronado,~E. {Molecular Spins for
  Quantum Computation}. \emph{Nat. Chem.} \textbf{2019}, \emph{11},
  301--309\relax
\mciteBstWouldAddEndPuncttrue
\mciteSetBstMidEndSepPunct{\mcitedefaultmidpunct}
{\mcitedefaultendpunct}{\mcitedefaultseppunct}\relax
\EndOfBibitem
\bibitem[Atzori and Sessoli(2019)Atzori, and Sessoli]{Atzori2019}
Atzori,~M.; Sessoli,~R. {The Second Quantum Revolution: Role and Challenges of
  Molecular Chemistry}. \emph{J. Am. Chem. Soc.} \textbf{2019}, \emph{141},
  11339--11352, PMID: 31287678\relax
\mciteBstWouldAddEndPuncttrue
\mciteSetBstMidEndSepPunct{\mcitedefaultmidpunct}
{\mcitedefaultendpunct}{\mcitedefaultseppunct}\relax
\EndOfBibitem
\bibitem[Khapaev \latin{et~al.}(2002)Khapaev, Kupriyanov, Goldobin, and
  Siegel]{Khapaev2002}
Khapaev,~M.~M.; Kupriyanov,~M.~Y.; Goldobin,~E.; Siegel,~M. {Current
  Distribution Simulation for Superconducting Multi-Layered Structures}.
  \emph{Supercond. Sci. Technol.} \textbf{2002}, \emph{16}, 24--27\relax
\mciteBstWouldAddEndPuncttrue
\mciteSetBstMidEndSepPunct{\mcitedefaultmidpunct}
{\mcitedefaultendpunct}{\mcitedefaultseppunct}\relax
\EndOfBibitem
\bibitem[Weil and Anderson(1965)Weil, and Anderson]{Weil1965}
Weil,~J.~A.; Anderson,~J.~K. {1039. The Determination and Reaction of
  2{,}2-Diphenyl-1-Picrylhydrazyl with Thiosalicylic Acid}. \emph{J. Chem.
  Soc.} \textbf{1965}, 5567--5570\relax
\mciteBstWouldAddEndPuncttrue
\mciteSetBstMidEndSepPunct{\mcitedefaultmidpunct}
{\mcitedefaultendpunct}{\mcitedefaultseppunct}\relax
\EndOfBibitem
\bibitem[Yordanov(1996)]{Yordanov1996}
Yordanov,~N.~D. {Is our Knowledge about the Chemical and Physical Properties of
  DPPH Enough to Consider It as a Primary Standard for Quantitative EPR
  Spectrometry?} \emph{Appl. Magn. Reson.} \textbf{1996}, \emph{10},
  339--350\relax
\mciteBstWouldAddEndPuncttrue
\mciteSetBstMidEndSepPunct{\mcitedefaultmidpunct}
{\mcitedefaultendpunct}{\mcitedefaultseppunct}\relax
\EndOfBibitem
\bibitem[\v{Z}ili\'c \latin{et~al.}(2010)\v{Z}ili\'c, Paji\'c, Juri\'c,
  Mol\v{c}anov, Rakvin, Planini\'c, and Zadro]{Zilic2010}
\v{Z}ili\'c,~D.; Paji\'c,~D.; Juri\'c,~M.; Mol\v{c}anov,~K.; Rakvin,~B.;
  Planini\'c,~P.; Zadro,~K. {Single Crystals of DPPH Grown from Diethyl Ether
  and Carbon Disulfide Solutions -- Crystal Structures, IR, EPR and
  Magnetization Studies}. \emph{J. Magn. Reson.} \textbf{2010}, \emph{207},
  34--41\relax
\mciteBstWouldAddEndPuncttrue
\mciteSetBstMidEndSepPunct{\mcitedefaultmidpunct}
{\mcitedefaultendpunct}{\mcitedefaultseppunct}\relax
\EndOfBibitem
\bibitem[Anderson and Weiss(1953)Anderson, and Weiss]{Anderson1953}
Anderson,~P.~W.; Weiss,~P.~R. {Exchange Narrowing in Paramagnetic Resonance}.
  \emph{Rev. Mod. Phys.} \textbf{1953}, \emph{25}, 269--276\relax
\mciteBstWouldAddEndPuncttrue
\mciteSetBstMidEndSepPunct{\mcitedefaultmidpunct}
{\mcitedefaultendpunct}{\mcitedefaultseppunct}\relax
\EndOfBibitem
\bibitem[H{\"o}cherl and Wolf(1965)H{\"o}cherl, and Wolf]{Hocherl1965}
H{\"o}cherl,~G.; Wolf,~H.~C. {Zur Konzentrationsabh{\"a}ngigkeit der
  Elektronenspin-Relaxationszeiten von Diphenyl-Picryl-Hydrazyl in Fester
  Phase}. \emph{Z. Phys.} \textbf{1965}, \emph{183}, 341--351\relax
\mciteBstWouldAddEndPuncttrue
\mciteSetBstMidEndSepPunct{\mcitedefaultmidpunct}
{\mcitedefaultendpunct}{\mcitedefaultseppunct}\relax
\EndOfBibitem
\bibitem[Piner \latin{et~al.}(1999)Piner, Zhu, Xu, Hong, and Mirkin]{Piner1999}
Piner,~R.~D.; Zhu,~J.; Xu,~F.; Hong,~S.; Mirkin,~C.~A. {"Dip-Pen"
  Nanolithography}. \emph{Science} \textbf{1999}, \emph{283}, 661--663\relax
\mciteBstWouldAddEndPuncttrue
\mciteSetBstMidEndSepPunct{\mcitedefaultmidpunct}
{\mcitedefaultendpunct}{\mcitedefaultseppunct}\relax
\EndOfBibitem
\bibitem[Bellido \latin{et~al.}(2010)Bellido, de~Miguel, Ruiz-Molina, Lostao,
  and Maspoch]{Bellido2010}
Bellido,~E.; de~Miguel,~R.; Ruiz-Molina,~D.; Lostao,~A.; Maspoch,~D.
  Controlling the Number of Proteins with Dip-Pen Nanolithography. \emph{Adv.
  Mater.} \textbf{2010}, \emph{22}\relax
\mciteBstWouldAddEndPuncttrue
\mciteSetBstMidEndSepPunct{\mcitedefaultmidpunct}
{\mcitedefaultendpunct}{\mcitedefaultseppunct}\relax
\EndOfBibitem
\bibitem[Mart\'inez-P\'erez \latin{et~al.}(2011)Mart\'inez-P\'erez, Bellido,
  Miguel, Ses\'e, Lostao, G\'omez-Moreno, Drung, Schurig, Ruiz-Molina, and
  Luis]{Martinez2011}
Mart\'inez-P\'erez,~M.~J.; Bellido,~E.; Miguel,~R.~d.; Ses\'e,~J.; Lostao,~A.;
  G\'omez-Moreno,~C.; Drung,~D.; Schurig,~T.; Ruiz-Molina,~D.; Luis,~F.
  {Alternating Current Magnetic Susceptibility of a Molecular Magnet
  Submonolayer Directly Patterned onto a Micro Superconducting Quantum
  Interference Device}. \emph{Appl. Phys. Lett.} \textbf{2011}, \emph{99},
  032504\relax
\mciteBstWouldAddEndPuncttrue
\mciteSetBstMidEndSepPunct{\mcitedefaultmidpunct}
{\mcitedefaultendpunct}{\mcitedefaultseppunct}\relax
\EndOfBibitem
\bibitem[Bellido \latin{et~al.}(2013)Bellido, Gonz\'alez-Monje, Repoll\'es,
  Jenkins, Ses\'e, Drung, Schurig, Awaga, Luis, and Ruiz-Molina]{Bellido2013}
Bellido,~E.; Gonz\'alez-Monje,~P.; Repoll\'es,~A.; Jenkins,~M.; Ses\'e,~J.;
  Drung,~D.; Schurig,~T.; Awaga,~K.; Luis,~F.; Ruiz-Molina,~D. {Mn$_{12}$
  Single Molecule Magnets Deposited on $\mu$-SQUID Sensors: the Role of
  Interphases and Structural Modifications}. \emph{Nanoscale} \textbf{2013},
  \emph{5}, 12565--12573\relax
\mciteBstWouldAddEndPuncttrue
\mciteSetBstMidEndSepPunct{\mcitedefaultmidpunct}
{\mcitedefaultendpunct}{\mcitedefaultseppunct}\relax
\EndOfBibitem
\bibitem[Bushev \latin{et~al.}(2011)Bushev, Feofanov, Rotzinger, Protopopov,
  Cole, Wilson, Fischer, Lukashenko, and Ustinov]{Bushev2011}
Bushev,~P.; Feofanov,~A.~K.; Rotzinger,~H.; Protopopov,~I.; Cole,~J.~H.;
  Wilson,~C.~M.; Fischer,~G.; Lukashenko,~A.; Ustinov,~A.~V. {Ultralow-Power
  Spectroscopy of a Rare-Earth Spin Ensemble Using a Superconducting
  Resonator}. \emph{Phys. Rev. B} \textbf{2011}, \emph{84}, 060501\relax
\mciteBstWouldAddEndPuncttrue
\mciteSetBstMidEndSepPunct{\mcitedefaultmidpunct}
{\mcitedefaultendpunct}{\mcitedefaultseppunct}\relax
\EndOfBibitem
\bibitem[Tavis and Cummings(1968)Tavis, and Cummings]{Tavis1968}
Tavis,~M.; Cummings,~F.~W. Exact Solution for an N-Molecule-Radiation-Field
  Hamiltonian. \emph{Phys. Rev.} \textbf{1968}, \emph{170}\relax
\mciteBstWouldAddEndPuncttrue
\mciteSetBstMidEndSepPunct{\mcitedefaultmidpunct}
{\mcitedefaultendpunct}{\mcitedefaultseppunct}\relax
\EndOfBibitem
\bibitem[H\"{u}mmer \latin{et~al.}(2012)H\"{u}mmer, Reuther, H\"{a}nggi, and
  Zueco]{Hummer2012}
H\"{u}mmer,~T.; Reuther,~G.~M.; H\"{a}nggi,~P.; Zueco,~D. Nonequilibrium Phases
  in Hybrid Arrays with Flux Qubits and Nitrogen-Vacancy Centers. \emph{Phys.
  Rev. A} \textbf{2012}, \emph{85}\relax
\mciteBstWouldAddEndPuncttrue
\mciteSetBstMidEndSepPunct{\mcitedefaultmidpunct}
{\mcitedefaultendpunct}{\mcitedefaultseppunct}\relax
\EndOfBibitem
\bibitem[Mart\'inez-P\'erez and Zueco(2019)Mart\'inez-P\'erez, and
  Zueco]{Martinez2019}
Mart\'inez-P\'erez,~M.~J.; Zueco,~D. {Strong Coupling of a Single Photon to a
  Magnetic Vortex}. \emph{ACS Photonics} \textbf{2019}, \emph{6},
  360--367\relax
\mciteBstWouldAddEndPuncttrue
\mciteSetBstMidEndSepPunct{\mcitedefaultmidpunct}
{\mcitedefaultendpunct}{\mcitedefaultseppunct}\relax
\EndOfBibitem
\bibitem[Ardavan \latin{et~al.}(2007)Ardavan, Rival, Morton, Blundell,
  Tyryshkin, Timco, and Winpenny]{Ardavan2007}
Ardavan,~A.; Rival,~O.; Morton,~J. J.~L.; Blundell,~S.~J.; Tyryshkin,~A.~M.;
  Timco,~G.~A.; Winpenny,~R. E.~P. {Will Spin-Relaxation Times in Molecular
  Magnets Permit Quantum Information Processing?} \emph{Phys. Rev. Lett.}
  \textbf{2007}, \emph{98}, 057201\relax
\mciteBstWouldAddEndPuncttrue
\mciteSetBstMidEndSepPunct{\mcitedefaultmidpunct}
{\mcitedefaultendpunct}{\mcitedefaultseppunct}\relax
\EndOfBibitem
\bibitem[Wedge \latin{et~al.}(2012)Wedge, Timco, Spielberg, George, Tuna,
  Rigby, McInnes, Winpenny, Blundell, and Ardavan]{Wedge2012}
Wedge,~C.~J.; Timco,~G.~A.; Spielberg,~E.~T.; George,~R.~E.; Tuna,~F.;
  Rigby,~S.; McInnes,~E. J.~L.; Winpenny,~R. E.~P.; Blundell,~S.~J.;
  Ardavan,~A. {Chemical Engineering of Molecular Qubits}. \emph{Phys. Rev.
  Lett.} \textbf{2012}, \emph{108}, 107204\relax
\mciteBstWouldAddEndPuncttrue
\mciteSetBstMidEndSepPunct{\mcitedefaultmidpunct}
{\mcitedefaultendpunct}{\mcitedefaultseppunct}\relax
\EndOfBibitem
\bibitem[Bader \latin{et~al.}(2014)Bader, Dengler, Lenz, Endeward, Jiang,
  Neugebauer, and van Slageren]{Bader2014}
Bader,~K.; Dengler,~D.; Lenz,~S.; Endeward,~B.; Jiang,~S.-D.; Neugebauer,~P.;
  van Slageren,~J. {Room Temperature Quantum Coherence in a Potential Molecular
  Qubit}. \emph{Nat. Commun.} \textbf{2014}, \emph{5}, 5304\relax
\mciteBstWouldAddEndPuncttrue
\mciteSetBstMidEndSepPunct{\mcitedefaultmidpunct}
{\mcitedefaultendpunct}{\mcitedefaultseppunct}\relax
\EndOfBibitem
\bibitem[Shiddiq \latin{et~al.}(2016)Shiddiq, Komijani, Duan, Gaita-Ari{\~n}o,
  Coronado, and Hill]{Shiddiq2016}
Shiddiq,~M.; Komijani,~D.; Duan,~Y.; Gaita-Ari{\~n}o,~A.; Coronado,~E.;
  Hill,~S. {Enhancing Coherence in Molecular Spin Qubits {\em via} Atomic Clock
  Transitions}. \emph{Nature} \textbf{2016}, \emph{531}, 348--351\relax
\mciteBstWouldAddEndPuncttrue
\mciteSetBstMidEndSepPunct{\mcitedefaultmidpunct}
{\mcitedefaultendpunct}{\mcitedefaultseppunct}\relax
\EndOfBibitem
\bibitem[Zadrozny \latin{et~al.}(2015)Zadrozny, Niklas, Poluektov, and
  Freedman]{Zadrozny2015}
Zadrozny,~J.~M.; Niklas,~J.; Poluektov,~O.~G.; Freedman,~D.~E. {Millisecond
  Coherence Time in a Tunable Molecular Electronic Spin Qubit}. \emph{ACS Cent.
  Sci.} \textbf{2015}, \emph{1}, 488--492\relax
\mciteBstWouldAddEndPuncttrue
\mciteSetBstMidEndSepPunct{\mcitedefaultmidpunct}
{\mcitedefaultendpunct}{\mcitedefaultseppunct}\relax
\EndOfBibitem
\bibitem[Ghirri \latin{et~al.}(2016)Ghirri, Bonizzoni, Troiani, Buccheri,
  Beverina, Cassinese, and Affronte]{Ghirri2016}
Ghirri,~A.; Bonizzoni,~C.; Troiani,~F.; Buccheri,~N.; Beverina,~L.;
  Cassinese,~A.; Affronte,~M. {Coherently Coupling Distinct Spin Ensembles
  through a High-${T}_{c}$ Superconducting Resonator}. \emph{Phys. Rev. A}
  \textbf{2016}, \emph{93}, 063855\relax
\mciteBstWouldAddEndPuncttrue
\mciteSetBstMidEndSepPunct{\mcitedefaultmidpunct}
{\mcitedefaultendpunct}{\mcitedefaultseppunct}\relax
\EndOfBibitem
\bibitem[Mergenthaler \latin{et~al.}(2017)Mergenthaler, Liu, Le~Roy, Ares,
  Thompson, Bogani, Luis, Blundell, Lancaster, Ardavan, Briggs, Leek, and
  Laird]{Mergenthaler2017}
Mergenthaler,~M.; Liu,~J.; Le~Roy,~J.~J.; Ares,~N.; Thompson,~A.~L.;
  Bogani,~L.; Luis,~F.; Blundell,~S.~J.; Lancaster,~T.; Ardavan,~A.; Briggs,~G.
  A.~D.; Leek,~P.~J.; Laird,~E.~A. {Strong Coupling of Microwave Photons to
  Antiferromagnetic Fluctuations in an Organic Magnet}. \emph{Phys. Rev. Lett.}
  \textbf{2017}, \emph{119}, 147701\relax
\mciteBstWouldAddEndPuncttrue
\mciteSetBstMidEndSepPunct{\mcitedefaultmidpunct}
{\mcitedefaultendpunct}{\mcitedefaultseppunct}\relax
\EndOfBibitem
\bibitem[Bonizzoni \latin{et~al.}(2017)Bonizzoni, Ghirri, Atzori, Sorace,
  Sessoli, and Affronte]{Bonizzoni2017}
Bonizzoni,~C.; Ghirri,~A.; Atzori,~M.; Sorace,~L.; Sessoli,~R.; Affronte,~M.
  {Coherent Coupling between Vanadyl Phthalocyanine Spin Ensemble and Microwave
  Photons: towards Integration of Molecular Spin Qubits into Quantum Circuits}.
  \emph{Sci. Rep.} \textbf{2017}, \emph{7}, 13096\relax
\mciteBstWouldAddEndPuncttrue
\mciteSetBstMidEndSepPunct{\mcitedefaultmidpunct}
{\mcitedefaultendpunct}{\mcitedefaultseppunct}\relax
\EndOfBibitem
\bibitem[Horcas \latin{et~al.}(2007)Horcas, Fern\'andez,
  G\'omez-Rodr\'{\i}guez, Colchero, G\'omez-Herrero, and Baro]{Horcas2007}
Horcas,~I.; Fern\'andez,~R.; G\'omez-Rodr\'{\i}guez,~J.~M.; Colchero,~J.;
  G\'omez-Herrero,~J.; Baro,~A.~M. WSXM: A Software for Scanning Probe
  Microscopy and a Tool for Nanotechnology. \emph{Rev. Sci. Instrum.}
  \textbf{2007}, \emph{78}\relax
\mciteBstWouldAddEndPuncttrue
\mciteSetBstMidEndSepPunct{\mcitedefaultmidpunct}
{\mcitedefaultendpunct}{\mcitedefaultseppunct}\relax
\EndOfBibitem
\bibitem[Ne\v{c}as and Klapetek(2011)Ne\v{c}as, and Klapetek]{Necas2011}
Ne\v{c}as,~D.; Klapetek,~P. Gwyddion: an Open-Source Software for SPM Data
  Analysis. \emph{Centr. Eur. J. Phys.} \textbf{2011}, \emph{10}\relax
\mciteBstWouldAddEndPuncttrue
\mciteSetBstMidEndSepPunct{\mcitedefaultmidpunct}
{\mcitedefaultendpunct}{\mcitedefaultseppunct}\relax
\EndOfBibitem
\end{mcitethebibliography}

\end{document}